\documentstyle[aas2pp4,twoside]{article}

\righthead{}
\lefthead{WOLLACK, ET AL.}
\begin{document}                

%
%
%
\title{An Instrument For Investigation\\
of the Cosmic Microwave Background Radiation\\
at Intermediate Angular Scales\\}

\author{E.~J.~Wollack\altaffilmark{1}, M.~J.~Devlin, %
N.~Jarosik,\\ C.~B.~Netterfield, L.~Page, and D.~Wilkinson}

\affil{Princeton University, Department of Physics, Jadwin Hall,
Princeton, NJ 08544} 

\altaffiltext{1}{Present address:  National Radio Astronomy
Observatory, 2015 Ivy Road, Charlottesville, VA~22903.}

%
%
\begin{abstract}

We describe an off-axis microwave telescope for observations of the
anisotropy in the cosmic microwave background (CMB) radiation on
angular scales between $0.5^{\circ}$ and $3^{\circ}$. The receiver
utilizes cryogenic high-electron-mobility transistor (HEMT) amplifiers
and detects the total power in multiple 3\,GHz wide channels. Both
frequency and polarization information are recorded allowing
discrimination between CMB radiation and potential foreground sources
and allowing checks for systematic effects.  The instrumental radiometric
offset is small ($\sim 1~$mK).  Data are taken by rapidly sampling
while sweeping the beam many beamwidths across the sky. After
detection, a spatio-temporal filter is formed in software which
optimizes the sensitivity in a multipole band in the presence of
atmospheric fluctuations. Observations were made from Saskatoon,
Saskatchewan (SK), Canada during the winter of 1993 with six channels
between 27.6 and 34.0\,GHz, in 1994 with twelve channels between 27.6
and 44.1\,GHz, and in 1995 with six channels between 38.2 and
44.1\,GHz. The performance of the instrument and assessment of the
atmospheric noise at this site are discussed.

\end{abstract}

\keywords{cosmic microwave background --- cosmology: atmospheric
effects: instrumentation --- radiometry --- telescopes}

\clearpage                      
\pagestyle{myheadings}

%
%
\section{Introduction}
\label{section:introduction}

In most cosmological models, the anisotropy in the cosmic microwave
background (CMB) is a direct probe of the conditions in the early
universe. A complete characterization of the $\sim 30~\mu$K signal can
potentially tell us about the large scale geometry of the universe, the
Hubble constant, the source of primordial density fluctuations, the
fraction of the universe made of baryons, and the ionization history
(\cite{bond94}, \cite{jungman95}).  Measurements of the CMB anisotropy
over a wide range of frequencies and angular scales are currently being
conducted. Reviews of the results and theoretical issues may be found in
\cite{bond95}, \cite{readhead92}, and \cite{white94}. In this paper, we
describe a ground-based telescope that operates between 28 and 45 GHz
and probes angular scales between 0.5 and 3 degrees (roughly between
multipoles $l = 60$ and 360). Data have been taken with both ${\rm
  K_a}$-band (28-34 GHz) and Q-band (38-45 GHz) receivers. Observations
were made in the winters of 1993, 1994, and 1995 at an observing site in
Saskatoon, Saskatchewan (SK), Canada. We refer to the results from the
three years as SK93, SK94, and SK95. Previous results and limited
information about the instrument are given in \cite{wollack93} (SK93),
\cite{wollack94a} (SK93), \cite{wollack94b}, \cite{page94},
\cite{netterfield95a} (SK94), and \cite{netterfield95b} (SK93 to SK94
comparison). An analysis of the data from all three years is in
\cite{netterfield96}.

\begin{figure}[tbph]
\plotone{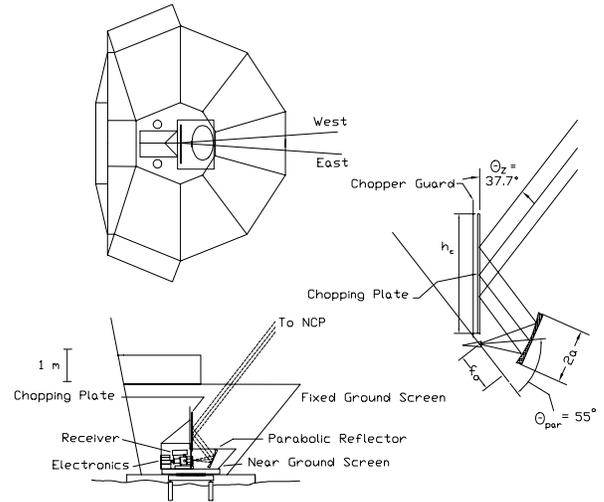}
\caption{The observing setup in Saskatoon and the telescope
  geometry. On left top and bottom are plan and section views of the
  SK94 telescope and ground-screen. On the right is a detail of the beam
  forming optics.} 
\label{plot:setup}
\end{figure}

        Figure~\ref{plot:setup} shows the configuration of the SK
experiment. A mechanically-cooled HEMT-based
(high-electron-mobility transistor) receiver senses the sky in
three frequency bands and two polarizations.  The beam is formed
by a cryogenic feed and an ambient temperature off-axis parabola.
The beam is steered on the sky by a large computer-controlled
chopping plate that oscillates around a vertical at $\sim4\,$Hz. 
As the plate moves, the receiver outputs are rapidly sampled. 
The elevation of the beam is fixed at $52^{\circ}$, the latitude
of Saskatoon, SK. To point the telescope, the receiver, parabola,
near ground-screen, electronics, and chopping plate are moved in
azimuth as a unit. All of the beam-forming optics are inside a large
fixed aluminum ground-screen. 

        For all measurements of the anisotropy of the CMB, the signal
is between $10^{-6}$ and $10^{-8}$ of the ambient temperature. Because
it takes anywhere from a few hours to hundreds of hours of integration
to distinguish the celestial signal from the atmospheric emission and
instrument noise, great care must be taken to ensure that the signal
is fixed to the sky and not due to a systematic effect buried in the
instrument noise. Generally, systematic effects modulate the baseline
or `offset' of the radiometer. The offset is the instrumental
contribution to a differential measurement of a uniform sky. In
principle this is zero. In practice, it typically ranges between tens
of micro-Kelvin to a few milli-Kelvin. It may be caused by atmospheric
gradients, thermal or emissivity gradients in the optics, polarized
emission from the optics, misalignment of the optics, asymmetric
sidelobe contamination, microphonics, or electronic pickup. In most
all experiments (one exception is the OVRO RING experiment
\cite{myers93}), this term is subtracted from the data before
analysis. The larger the offset, the more vulnerable the data are to
the environment and quirks of the instrument. Just as important as the
magnitude is the offset's stability.

        Since the discovery of the CMB, researchers have realized that
the radiometric offset can be minimized by moving only the optical
element furthest from the receiver (\cite{partridge67},
\cite{fabbri80}, \cite{lubin83}, \cite{davies87}, \cite{dalloglio88},
\cite{dragovan94}). If the beam is scanned only in azimuth, then the
atmospheric signal is minimized.  The remaining dominant sources are
modulated emission from and spill past the edges of the chopping
plate. For the Saskatoon experiments, the offset depends on the year
and the observing strategy. In 1993 and 1994, it was $\sim 400~\mu$K;
in the worst case in 1995, it was $2.2~$mK. The change in average
offset was $<7~\mu$K$/$day for all observations.

        For clarity, this paper focuses on the ${\rm
K_a}$-band system, but frequent references are  made to the two
different Q-band configurations that share the same observing
platform. In Section~\ref{section:radiometer} the receiver is
discussed. In Section~\ref{section:optics} we outline the telescope
design and performance. Calibration, radiometric offsets, and
atmospheric seeing are reviewed in
Sections~\ref{section:calibration}, ~\ref{section:offset}, and
\ref{section:atmos} respectively.

%
\section{The Receiver} 
\label{section:radiometer} 
\begin{figure}[tbph]
\plotone{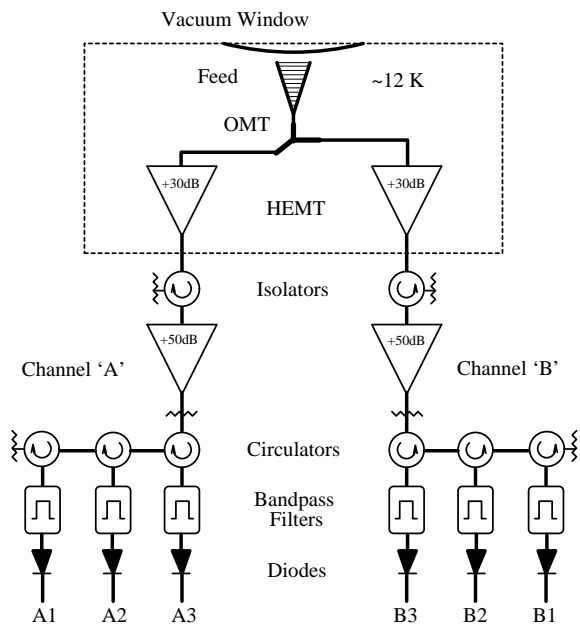}
\caption{The ${\rm SK93K_a}$ receiver
  layout. Radiation enters through 
  the vacuum window and is collected by a cooled corrugated feed. It is
  then split into two orthogonal polarizations by the orthomode
  transducer (OMT). Each polarization is then amplified, separated into
  three frequency bands, and square-law detected.} 
\label{plot:radiometer}
\end{figure}

        The receiver is comprised of a mechanically-cooled dewar at
15\,K, which houses the HEMT amplifiers and a single feed horn, and a
thermally regulated 300\,K enclosure, which houses the warm amplifiers
and band-defining microwave components.  Figure~\ref{plot:radiometer}
shows a schematic of the ${\rm K_a}$ receiver. It operates in
total-power, base band mode; that is, there are no Dicke switches, mixers,
or local-oscillators. Similar receivers are discussed in
\cite{gaier92} and \cite{gundersen94}. The receiver is positioned so
that the feed horn's phase center is imaged to within 0.5\,cm of the
rotation axis of the chopper which is in turn near the center of a
large fixed ground-screen. The geometry is shown in
Figure~\ref{plot:setup}.

\subsection{Receiver RF Design}  
\label{subsection:rec_rf}

\begin{deluxetable}{lccc}
\small
\tablewidth{0pc}
\tablecaption{SK Telescope Specifications \label{table:telescope}}
\tablehead{     & \colhead{SK93${\rm K_a}$}
                & \colhead{SK94${\rm K_a/Q}$}
                & \colhead{SK95${\rm Q}$}}

\startdata 
Feed Semi-flare Angle, $\theta_{\rm o}$                                
        & $6.0^\circ$           
        & $6.0^\circ$   
        & $4.4^\circ$   \nl
Feed Aperture Diameter, $d_{\rm h}$                             
        & $4.2\,$cm             
        & $4.2/3.0\,$cm 
        & $2.1\,$cm     \nl
Feed Normalized Phase Error, $\Delta$                           
        & $<0.1$                
        & $<0.1$        
        & $<0.06$       \nl
Feed Phase Center, $t(\nu_{\rm c})$  
        & $1.5\,$cm             
        & $1.5\,$cm/$1.0\,$cm           
        & $0.24\,$cm    \nl
Feed Hybrid Frequency, $\nu_{\rm x}$                            
        & $29.9\,$GHz           
        & $29.9/39.7\,$GHz      
        & $39.7\,$GHz   \nl
Feed Beamwidth, $\theta^{\rm FWHM}_{\rm feed}(\nu_{\rm x})$
        & $18^\circ$            
        & $18^\circ/20^\circ$   
        & $27^\circ$    \nl
Primary Edge Illumination                                       
        & $<-24\,$dB            
        & $<-24\,$dB    
        & $<-27\,$dB    \nl \hline
Primary Offset Angle, $\theta_{\rm par}$                         
        & $+55.0^\circ$         
        & $+55.0^\circ$ 
        & $+55.0^\circ$ \nl
Primary Minor Axis, $2b$                                        
        & $58.4\,$cm            
        & $58.4\,$cm    
        & $122.0\,$cm   \nl
Primary Major Axis, $2a$                                        
        & $65.8\,$cm            
        & $65.8\,$cm    
        & $137.5\,$cm   \nl
Primary Focal Length, $f_{\rm o}$                                     
        & $50.0\,$cm            
        & $50.0\,$cm    
        & $75.0\,$cm    \nl
Primary Surface Tolerance, $\Delta_{\rm rms}$                   
        & $8\mu$m               
        & $8\mu$m       
        & $\sim 10-20\mu$m      \nl
Primary Aperture Efficiency, $\eta_{\rm a}(\nu_{\rm c})$                   
        & $0.52$                
        & $0.54$        
        & $0.56$        \nl
Primary Beamwidth on Sky, $\theta^{\rm FWHM}_{\rm beam}(\nu_{\rm c})$     
        & $1.44^\circ$          
        & $1.42^\circ/1.04^\circ$       
        & $0.5^\circ$   \nl
Chopper Edge Illumination                                       
        & $<-45\,$dB    
        & $<-45\,$dB    
        & $<-30\,$dB    \nl
Chopper Shield Edge Illumination                                
        & $<-65\,$dB            
        & $<-65\,$dB    
        & $<-50\,$dB    \nl
Far Ground-screen Edge Illumination                             
        & $<-65\,$dB            
        & $<-65\,$dB    
        & $<-65\,$dB    \nl \hline
Chopper Width, $w_{\rm c}$                                            
        & $91.4\,$cm            
        & $91.4\,$cm    
        & $146.1\,$cm   \nl
Chopper Height, $h_{\rm c}$                                           
        & $121.9\,$cm           
        & $121.9\,$cm   
        & $207.6\,$cm   \nl
Chopper Surface Tolerance, $\Delta_{\rm rms}$                   
        & $<40\mu$m             
        & $<40\mu$m     
        & $<40\mu$m     \nl
Chopper Aperture Efficiency, $\eta_{\rm a}(\nu_{\rm c})$                   
        & $0.16$                
        & $0.17$        
        & $0.27$        \nl
Chopper Step Response Transition Time                           
        & $29\,$ms              
        & $32\,$ms      
        & $-$   \nl
Chopper Frequency, $f_{\rm c}\,$(max)                                 
        & $10\,$Hz      
        & $6\,$Hz       
        & $3\,$Hz       \nl
Chopper Amplitude in AZ, $\phi_{\rm c}\,$(max)                        
        & $\pm2.5^\circ$        
        & $\pm3.5^\circ$ 
        & $\pm3.5^\circ$ \nl 
\enddata
\end{deluxetable}

        Radiation from the sky enters the receiver through a 9\,cm
diameter vacuum window made of 0.38\,mm polypropylene.  Two
overlapping aluminum baffles define the entrance aperture. One is
anchored to the $\sim70\,$K stage, the other to ambient temperature.
Strips of aluminized Mylar electrically connect the top of the feed to
the warm baffle and suppress RF interference. To inhibit the formation
of frost on the vacuum window, warm air is forced into a volume in
front of the window defined by a low loss polystyrene foam ring
(Eccofoam PS, Emerson \& Cuming Inc., Canton, MA) covered with a
$0.09\,$mm polyethylene sheet.

        The beam is formed by a conical corrugated scalar feed (See
\cite{clarricoats84} and references therein.). At the base of the
feed, a wide band TE$^\circ_{11}$-to-HE$_{11}$ mode converter
(\cite{james82a}, \cite{wollack94a}) matches the hybrid mode to
circular waveguide. Next, for the ${\rm K_a}$ system, an electroformed
adiabatic round-to-square transition matches the feed to the orthomode
transducer (OMT) which splits the incident radiation into vertical and
horizonal linear polarizations.  The low end of the feed's bandpass is
defined by the input waveguide cutoff and the upper end is set by the
excitation of TM\,$^\circ_{11}$ in the horn-to-circular-guide
transition. Over the 26-to-36 GHz bandpass of the ${\rm K_a}$
receiver, the reflection coefficient is less than $-26\,$dB and
between 23 GHz and 48 GHz it is less than $-20\,$dB. The Q-band feed
for SK94 is scaled from the ${\rm K_a}$-band design and has similar
performance. The design parameters for all of the feeds are given in
Table~\ref{table:telescope}.

        The outputs of the OMT are fed directly into two National
Radio Astronomy Observatory (NRAO) amplifiers
(\cite{pospieszalski88,pospieszalski92}). These devices have
approximately $10\,$GHz of available bandwidth and $+30\,$dB of RF
gain. When cooled to $\sim15\,$K, the average noise temperatures of
the ${\rm K_a}$ and Q-band amplifiers are $\approx 50\,$K and $\approx
20\,$K respectively.

        The HEMT amplifiers are connected to ambient temperature
isolators by a $10\,$cm length of $0.25\,$mm wall stainless steel WR28
waveguide (WR22 for Q-band), gold plated to minimize attenuation. The
waveguide vacuum windows between the refrigerator and the room
temperature receiver box are made of $0.013\,$mm thick kapton sheet.
Following the isolators are commercial wide-band amplifiers with $\sim
+50 \,$dB of gain. These amplifiers have a noise figure of $4-7\,$dB.
Split-block attenuators are used to flatten the bandpass and set the
signal level presented to the diodes detectors in the ${\rm K_a}$
receiver. In the Q-band receiver, a relatively large HEMT gain slope
and the frequency multiplexor design necessitated locating the level
set attenuator before the room temperature amplifiers to avoid gain
compression.  In both receivers, the contribution to the total system
temperature from the components after the HEMT amplifiers is $<2\,$K.

\begin{deluxetable}{llll}
\footnotesize
\tablewidth{0pc}
\tablecaption{Radiometer Components \label{table:receiver}}
\tablehead{     & \colhead{Component} 
                & \colhead{Part Number} 
                & \colhead{Source} }

\startdata 
SK--${\rm K_a}$ & Orthomode Transducer      
                & P/N OM3800  
                & Atlantic Microwave Corp., Bolton, MA \nl
                & Cryogenic Amplifiers   
                & S/N 6, 14 
                & NRAO, Charlottesville, VA \nl
                & Circulators 
                & P/N WFR-C 
                & Microwave Resources Inc., Chino, CA\nl
                & Warm Amplifiers       
                & P/N SMW92-1953 
                & Avantek Inc., Santa Clara, CA\nl
                & Bandpass Filters `A' 
                & P/N BFF-28-27.5, 30.5, 33.5   
                & Dorado Int. Corp., Seattle, WA\nl
                & Bandpass Filters `B'
                & P/N F27.5-6, 30.5-6, 34.0-6   
                & Spacek Labs, Santa Barbara, CA\nl
                & Diodes         
                & P/N DXP-28    
                & Millitech Corp., South Deerfield, MA \nl 
\hline
SK--${\rm Q}$   & Orthomode Transducer  
                & P/N 110265-1  
                & Gamma --$f$ Corp., Torrance, CA\nl 
                & Cryogenic Amplifiers  
                & S/N B-32, B-33 
                & NRAO, Charlottesville, VA\nl
                & Isolators
                & P/N 45162H-1000
                & Hughes, Microwave Prod. Div., Torrance, CA\nl
                & Warm Amplifiers               
                & P/N DB93-0474 
                & DBS Microwave, El Dorado Hills, CA \nl
                & Triplexers            
                & P/N 3647                      
                & Pacific Millimeter Products, Golden, CO\nl
                & Diodes                        
                & P/N DXP-22 
                & Millitech Corp., South Deerfield, MA\nl
\enddata
\end{deluxetable}

        In the ${\rm K_a}$ receiver, half-wave waveguide filters and
full-band circulators form a filter bank as shown in
Figure~\ref{plot:radiometer}. The $\sim 10~$GHz HEMT bandwidth is
divided into three passbands in order to maximize spectral
discrimination, passband flatness, and minimize noise. The highest
usable frequency is set by the OMT. In the E-plane port (called A
channels), spurious spikes due to the ${\rm TM}_{11}^\Box$ mode in the
OMT set the upper band edge at $35.6\,$GHz. The passband of the
H-plane port (called B channels) smoothly degrades above $36\,$GHz.
The highest frequency filters are tailored for maximum bandwidth in
response to this difference. Some care must be taken with the lowest
channel filter. Waveguide dispersion moves the `second harmonic' of
the filter response to lower frequencies than is expected in a TEM
structure (\cite{matthaei80}). A `picket fence' filter with a
26-to-$32\,$ GHz bandpass in series with the filter is necessary to
prevent leakage near 38 GHz into the high frequency channels.  The
large number of waveguide joints in the filter bank make them
especially prone to microphonics. This problem was overcome by
embedding them in a mechanically stable and electrically lossy
contoured support structure. In the Q-band system, all the filtering
is done with a custom built, monolithic, channel-dropping
multiplexor. A listing of the major components used in the two
receivers is given in Table~\ref{table:receiver}.

        On the output of each waveguide filter is a negative polarity
Schottky barrier diode. A typical sensitivity is $1500~$mV/mW and the
response is linear in power for input levels small compared to
$-10\,$dBm (with a 1~M$\Omega$ audio impedance).  Optimal performance
was achieved with $-18\,$dBm of RF power at the diode. By tuning each
chain's split-block attenuator, all channels were within $3\,$dB of
this value. The typical DC level out of a diode was $\sim 20\,$mV.
Mounted on the SMA output of each diode is a low-noise preamp with a
gain of 100 and a $2{\rm k}\Omega$ input impedance. This impedance is
a reasonable compromise between the reduction in sensitivity
(compared to ${\rm 1\,M\Omega}$) and increased signal linearity and
thermal stability. The radiometer noise is approximately ten times the
preamp noise floor.

\subsection{Receiver Thermal and Mechanical Considerations}
\label{subsection:rec_thme}

        The cryogenic components are cooled by a CTI 350 refrigerator
which allows uninterrupted operation of the receiver for extended
periods of time.  All mechanical and electrical connections from the
cold stage to $300\,$K are thermally anchored to the refrigerator
$\sim70\,$K stage.  The bias wiring, waveguide connections to the room
temperature receiver, and radiation loading of the refrigerator result
in a total cold stage thermal load of $950~$mW. With a CTI 8500
compressor, the cold stage runs at $\sim12\,$K, roughly $2\,$K warmer
than without any thermal load. A {\rm rms} thermal variation of $\sim
40\,$mK synchronous with the refrigerator $1.2\,$Hz drive is measured
at the cold station. This signature is not detected in the radiometer
outputs.

        By design, the cold stage vibrational amplitude is
$<0.013\,$mm (\cite{cti}). The vibration power spectrum is
dominated by a $1.2\,$Hz line from the cold head cycle and by the
broadband feature at $\sim 200\,$Hz resulting from gas flow
through the refrigerator. Slight changes in RF impedance caused
by vibrations can be synchronously modulated by the chopper. 
This microphonic sensitivity was eliminated through the use of
monolithic sub-assemblies, adequate strain relief throughout the
receiver chain, and a waveguide flange alignment of $< 0.05\,$mm.
In particular, microphonics were traced to loose attenuator
vanes, isolator tuning networks, and waveguide joints.

        The room temperature receiver box is rigidly mounted on the
dewar.  It has two levels of RF shielding and is filled with microwave
absorber to limit air convection as well as damp microwave energy.  A
G-10 fiberglass spacer mechanically supports the inner box and
provides the thermal path between the boxes with a several hour time
constant.

        The diode sensitivity and amplifier gain are a function of
temperature. Double regulation of the temperature of the receiver
enclosure is required to allow field operation between $-45^\circ\,$C
and $0^\circ\,$C.  The outside of the enclosure is regulated to
$10\pm2^\circ\,$C , while the inside is regulated at $28\,$C to better
than $\pm0.035^\circ\,$C. In addition, the RF components are fitted
with insulation to prevent convective cooling and are thermally
anchored to a common aluminum mounting plate. With the inner regulator
on, the gain coefficient is less than $-0.03\,$dB/K for temperature
changes on the outside of the receiver box.

\subsection{Radiometer Noise and Passbands}
\label{subsection:rec_cal}

        We use two different means to measure the system noise in the
laboratory. In the first method, a variation on the Y-factor technique
(e.g. \cite{pozar90}), the power from a cold load is varied and the
radiometer response is recorded for a series of points. Over the
region of linear response, the sensitivity and noise are related to
the slope and the intercept of the resulting data.  This technique
measures the mean noise power. In the second method, the audio
frequency power spectrum of the calibrated receiver is measured
directly with a spectrum analyzer. Only at frequencies large compared
to the receiver $1/f$ knee do these two methods give the same spectral
noise density. The Y-factor test always yields a better sensitivity
because it is insensitive to low frequency fluctuations.

\begin{deluxetable}{lcccccc}
\small
\tablecaption{Radiometer Centroids, Passbands, and Sensitivity \label{table:noise}}
\tablehead{     & \colhead{$i$} 
                & \colhead{$\nu_{\rm c}^{ }$} 
                & \colhead{$\Delta\nu_{\rm rf}$} 
                & \colhead{$T_{\rm N}$$^{\rm a}$} 
                & \colhead{$S_{\rm lab}$$^{\rm b}$} 
                & \colhead{$S_{\rm sky}$$^{\rm c}$} \nl
                & \colhead{[Channel]}
                & \colhead{[GHz]}      
                & \colhead{[GHz]}      
                & \colhead{[K]}      
                & \colhead{[mK sec$^{1\over2}$]}
                & \colhead{[mK sec$^{1\over2}$]} }

\startdata 
SK93/94${\rm K_a}$ & ${\rm K_a}$--A1 & $27.6\pm0.15$ & $2.8\pm0.2$ 
                & $44$  & $1.1$ & $1.6$\nl
                & ${\rm K_a}$--A2 & $30.5\pm0.15$ & $2.9\pm0.2$ 
                & $38$  & $1.5$ & $1.9$\nl
                & ${\rm K_a}$--A3 & $33.4\pm0.15$ & $2.8\pm0.2$ 
                & $53$  & $1.3$ & $1.6$\nl
                & ${\rm K_a}$--B1 & $27.5\pm0.15$ & $2.8\pm0.2$ 
                & $29$  & $1.1$ & $1.5$\nl
                & ${\rm K_a}$--B2 & $30.5\pm0.15$ & $2.8\pm0.2$ 
                & $47$  & $1.3$ & $1.7$\nl
                & ${\rm K_a}$--B3 & $34.0\pm0.15$ & $3.7\pm0.2$ 
                & $51$  & $1.4$ & $1.7$\nl 
\hline

SK94Q          & ${\rm Q}$--B1   & $38.2\pm0.12$ & $2.5\pm0.1$ 
               & $<10$           & $1.8$ & $2.3$\nl
               & ${\rm Q}$--B2   & $40.7\pm0.12$ & $4.1\pm0.1$ 
               & $10$            &       & $2.5$\nl
               & ${\rm Q}$--B3   & $44.1\pm0.12$ & $3.3\pm0.1$  
               & $15$            &       & $4.0$\nl  
\hline

SK95Q           & ${\rm Q}$--A1   & $38.3\pm0.12$ & $2.6\pm0.1$ 
                & $21$  & $2.5$ & $3.5$ \nl
                & ${\rm Q}$--A2   & $40.8\pm0.12$ & $2.3\pm0.1$ 
                & $18$  & $1.9$ & $3.0$ \nl
                & ${\rm Q}$--A3   & $44.8\pm0.12$ & $1.8\pm0.1$ 
                & $24$  & $2.9$ & $5.1$ \nl 
                & ${\rm Q}$--B1   & $38.3\pm0.12$ & $1.8\pm0.1$ 
                & $25$  & $2.1$ & $2.8$ \nl
                & ${\rm Q}$--B2   & $41.1\pm0.12$ & $3.3\pm0.1$ 
                & $22$  & $1.3$ & $1.7$ \nl
                & ${\rm Q}$--B3   & $44.7\pm0.12$ & $2.5\pm0.1$ 
                & $16$  & $2.2$ & $3.3$ \nl  
\enddata
\tablecomments{(a) The reported receiver noise temperatures, $T_{\rm N}$,
are measured at the OMT circular waveguide input flange. The
uncertainty is $\pm 3\,$K. (b) Receiver noise (total power) measured
in the lab at 8 Hz using a $15\,$K cold load for ${\rm
K_a}$ and 20\,K for the Q-band system. The increase over the ideal is
due to imperfect passbands and `$1/f$' noise in the HEMT
amplifier. The error is $\pm 0.2\,$mK\,sec$^{1\over2}$.  (c) System
noise at $8\,$Hz while observing the sky. These results are for
atmospheric noise cut levels of $\zeta = 4.5$, 1.7, and $1.5\,{\rm
mK\,sec^{1\over2}\,deg^{-1}}$ for ${\rm SK93/94K_a}$, SK94Q, and
SK95Q. The sensitivity to {\rm CMB} fluctuations is reduced by a
factor of $\kappa_3$ due to the observing pattern lock-in (See
Table~\ref{table:kappa}).} 
\end{deluxetable}

\begin{figure}[tbph]
\plotone{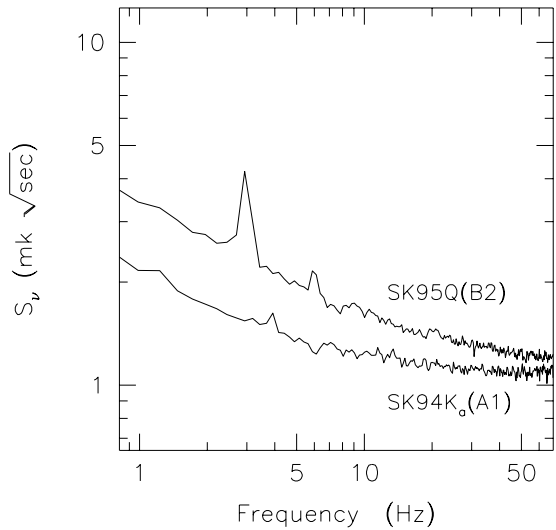}
\caption{The power spectra, $S_\nu$, of the receiver
  output. The data 
  were taken while observing the sky during good weather. For SK95Q, the
  2pt and 3pt offsets (See
  Sections~\protect{\ref{section:stratsynth}}
  and~\protect{\ref{section:offset}}) are manifest as the spectral
  features at 3 and 6\,Hz ($\zeta = 1.5\,{\rm
    mK\,sec^{1\over2}\,deg^{-1}}$). For ${\rm SK94K_a}$, only the
  2pt offset at 4\,Hz is evident ($\zeta = 0.9\,{\rm
    mK\,sec^{1\over2}\,deg^{-1}}$). Data taken under laboratory conditions
  have an indistinguishable $1/f$ component which originates from the
  cooled HEMT's. Given the approximately equal bandwidths in the ${\rm
    K_a}$ and Q systems, the higher Q-band $1/f$ knee reflects a lower
  gain stability. At $f\gg 100\,$Hz the power spectra for both
  radiometers agree with the sensitivity derived from the measured RF
  bandwidth and system temperature.} 
\label{plot:tpfft}
\end{figure}

        In the power spectrum of the noise at the diodes, as shown in
Figure~\ref{plot:tpfft}, the $1/f$ character is readily apparent. This
was shown to be due to gain fluctuations in the HEMT amplifiers
(\cite{jarosik93}). The total power spectral density is well modeled
by a modified radiometer equation (e.g. \cite{rohlfs90}),
\begin{equation}
{\Delta T\over T_{\rm sys}} =
\sqrt{{1\over\Delta \nu_{\rm rf} \tau}+\left({\Delta G(f)\over G}\right)^2},
\end{equation}
where $(\Delta G/G)^2 \propto 1/f^\alpha$ is the square of the
fractional receiver gain variation. The dominant contribution to the
gain fluctuations in the system is the cryogenic HEMT amplifier.  The
measured variations in the ${\rm K_a}$ band HEMT gain are, $\left(
{\Delta G / G} \right) \sim 2\times 10^{-5}$ at $8\,$Hz with
$\alpha\simeq 0.9$. Both the magnitude of the fluctuations and
$\alpha$ are weakly dependent on the transistor bias
(\cite{wollack95}). A summary of the measured sensitivities is given
in Table~\ref{table:noise}.

        The gain fluctuations correlate data from different
channels of a single amplifier chain. With the independent A and
B polarizations, which are not correlated by gain fluctuations,
one can distinguish between correlations due to a common source,
for instance a fluctuating atmosphere, and those intrinsic to
each amplifier. As the magnitude of the atmospheric fluctuations
approaches the receiver noise floor, the cross-polarization
frequency-frequency correlations ($<AB>$) tend to zero, whereas
the frequency-frequency correlations ($<AA>$, $<BB>$) approach
the same value as measured in the lab with a cold load. Any
inter-channel correlation, instrumental or atmospheric in origin,
must be accounted for in assessing the statistical significance
of the data (\cite{wollack93}, \cite{dodelson94}).

        The bandpass is determined by measuring the
power response in each band with a leveled Hewlett Packard 8690B RF
swept source at the orthomode transducer flange. It is defined as
\begin{equation}
\Delta \nu_{\rm rf} = {(\int_0^\infty G(\nu)
d\nu)^2 \over \int_0^\infty G(\nu)^2 d\nu },
\label{equation:dneff}
\end{equation}
where $G$ is the power response profile of the microwave band pass
(\cite{dicke46}). In a typical channel, the bandpass ripple results in
a $\sim 5\%$ reduction in sensitivity relative to an ideal filter
response. Similarly, the center frequency is computed from
\begin{equation}
\nu_{\rm c}^{ } = {\int_0^\infty \nu T_{\rm b}(\nu) G(\nu) 
d\nu \over \int_0^\infty T_{\rm b}(\nu) G(\nu) d\nu },
\end{equation}
where $T_{\rm b}$ is the brightness temperature of the calibration
source. The center frequency for a blackbody and a synchrotron
calibration source (e.g. Cas\,A) differ by $\sim200\,$MHz for a
typical channel. The effective bandwidths and center frequencies are
given in Table~\ref{table:noise}.

\subsection{The Data Collection System}
\label{subsection:datasystem}

\begin{deluxetable}{lccc}
\tablewidth{0pc}
\tablecaption{SK Observing Parameters \label{table:observe}}
\tablehead{     & \colhead{SK93${\rm K_a}$}
                & \colhead{SK94${\rm K_a/Q}$}
                & \colhead{SK95${\rm Q}$} }

\startdata 
Elevation Angle, $\theta_{\rm EL}$
        & $+52.25\pm0.06^\circ$ 
        & $+52.16\pm0.06^\circ$ 
        & $+52.24\pm0.01^\circ$ \nl
Primary Beamwidth, $\theta^{\rm FWHM}_{\rm beam}(\nu_{\rm c})$
        & $1.44\pm0.02^\circ$ 
        & $1.42\pm0.02^\circ$/$1.04\pm0.02^\circ$       
        & $0.5^\circ$   \nl
Point Source Sensitivity, $\Gamma (\nu_{\rm c})$
        & $49\,{\rm \mu K\,Jy^{-1}}$
        & $50\,{\rm \mu K\,Jy^{-1}}/ 52\,{\rm \mu K\,Jy^{-1}}$
        & $230\,{\rm \mu K\,Jy^{-1}}$   \nl
Chopper Frequency, $f_{\rm c}$
        & $3.906\,$Hz   
        & $3.906\,$Hz           
        & $2.970\,$Hz           \nl
Sample Frequency, $f_{\rm s}$
        & $62.5\,$Hz    
        & $250\,$Hz             
        & $500\,$Hz             \nl
Sky Detection Frequency
        & $12\,$Hz      
        & $8 - 32\,$Hz          
        & $6 - 70\,$Hz          \nl
Samples per Chop, $N_{\rm s}$
        & $16$          
        & $64$                  
        & $168$         \nl
Chopper Phase Offset, $\delta N_{\rm s}$
        & $+0.6\pm0.1$  
        & $0.0\pm0.1$   
        & $0.0\pm0.002$ \nl
Bessel Filter Roll-Off, $f_{\rm 3dB}$           
        & $30\,$Hz      
        & $120\,$Hz             
        & $250\,$Hz     \nl
Beams per Throw         
        & $4$
        & $5/7$         
        & $17$  \nl
Chopper Amplitude on Sky, $\theta{\rm_t}$       
        & $\pm2.45^\circ$ 
        & $\pm3.51^\circ$/$\pm3.68^\circ$       
        & $\pm3.97^\circ$       \nl
Base Wobble Amplitude on Sky, $\theta{\rm_w}$   
        & $\pm4.91^\circ$ 
        & $\pm4.42^\circ$       
        & $\pm4.50^\circ$       \nl
Observing Polarization, A B     
        & $\updownarrow\leftrightarrow$ 
        & $\updownarrow\leftrightarrow$/$\ldots \leftrightarrow$        
        & $\updownarrow\leftrightarrow$ \nl
\enddata
\end{deluxetable}

        A 80386-based computer located outside the fixed ground screen
controls the telescope and data acquisition. The six receiver channels
are read with V/F (voltage-to-frequency) converters on the
telescope. Counter cards on the PC backplane synchronously `integrate'
the V/F output for each sample (See Table~\ref{table:observe}).  To
have sufficient dynamic range, each signal's DC level is removed by a
low-noise bucking circuit and then amplified by a factor of five.  The
signals are filtered by eight-pole low-pass Bessel filters with a time
constant determined by the sample rate. The system has an adjustable
$\sim10\,$K data window. In practice, we find a $4\,$K window is
sufficient for all the usable sky data (See
Section~\ref{section:atmos}). The chopper and base position are
recorded at the same rate as the primary data.  Housekeeping data are
read by a 12 bit A/D in the PC at a slower rate. One master clock
controls the V/F converters, the counter gates, and the chopper/base
pointing update. In 1995, roughly 1\,GByte/day of data were
recorded. The telescope-control/data-acquisition PC is connected via
Ethernet to a workstation where the data are stored, reduced and
analyzed.

\subsection{Electrical Offsets and Interference}
\label{subsection:pickup}

        Synchronous electrical pick-up and crosstalk in the system are
small. A monitor diode and preamp in the receiver with a factor of
$1\times10^3$ higher audio gain than the other channels indicates that
synchronous electrical signals are less than $0.1~\mu$K. The receiver
was tested in the lab with a regulated cold load as the radiation
source while the chopper underwent a three position 4\,Hz chop (more
gentle motions were used in the field). This maximizes the chopper
drive current and vibration. After a $\sim 1\,$day coherent average of
the data, an absolute upper limit of $< 10~\mu$K is set on all audio
frequencies of interest on any potential electrical or microphonic
pick-up.  In the field, the signal from the quadrature
phase~\footnote{The quadrature phase, which is not sensitive to a
signal on the sky, is shifted $90^\circ$ relative to the chopper phase
which is sensitive to the celestial signal. See
Section~\ref{section:calibration}} of the weighted data are stable and
are typically $\sim 20 \pm 10 \mu\,$K; in worst case for SK95Q, $< 70
\mu\,$K.

        The Saskatoon airport radar operates in L-band ($\sim1\,$GHz)
and sweeps our site every $8\,$seconds. This signal was not detected
in the data or in the monitor channel at the ${\rm 20\,\mu K}$ level.
In the laboratory the receiver and data collection system were
illuminated with a + 12\,dBm L-band source, again no signal was seen
at the ${\rm 10\,\mu K}$ level. There are no indications of RF
interference in any of the data~\footnote{In the last decade the
population of satellites using ${\rm K_a}$-band communication channels
has increased dramatically. There are now approximately 70 satellites
in the Space Network Listing (\cite{rcb93}) with on board equipment
operating between 20 to $40$\,GHz. The bulk of these emit in our
lowest frequency channel from geosynchronous orbits. Given typical
satellite beam coverage, orbital parameters, and our scan strategy,
the likelihood of RF contamination from a communications satellite is
small.}.

%
%
%
\section{Antenna Design and Performance}
\label{section:optics}

        An offset-parabolic reflector fed by a corrugated feed
has minimal blockage, approximately equal E/H-plane beamwidths,
and relatively low sidelobe response (\cite{rudge82}). These
properties are ideal for CMB observations. The radiometer beam is
formed by a cooled corrugated feed horn which under-illuminates
an ambient temperature parabola which in turn illuminates the
chopping flat.  There are two levels of ground-screen shielding:
the near ground-screen moves with the beam-forming optics, 
shielding the telescope base and decking, and the stationary far
ground-screen shields the earth and sun. The telescope is
optimized for observations at the elevation of the North
Celestial Pole (NCP) from the site. See Figure~\ref{plot:setup}
and Table~\ref{table:telescope} for a summary of the telescope
geometry and system parameters.

\subsection{Primary Illumination and The Main Beam}

        When an under-illuminated parabolic section is fed with a
diffraction-limited feed, the resulting beam size is approximately
frequency independent. This occurs because the competing effects of the
diffraction-limited feed and the under-filled primary cancel. In this
limit, the effective diameter of the primary illumination is inversely
proportional to frequency.

        For a corrugated feed with a small aperture phase
error~\footnote{The aperture phase error is the difference in
wavelengths between the path from the feed apex to the edge of the
aperture and from the apex to the center of the
aperture~(See for example, \cite{thomas78}). It is defined as
\begin{equation}
\Delta \equiv {l_{\rm h} \over \lambda} \left(1-\cos{\theta_o}
\right)= {d_{\rm h}\over 2 \lambda} \tan\left({\theta_o \over
2}\right),
\end{equation}
where $l_{\rm h}$ is the horn slant length, $d_{\rm h}$ is the
aperture diameter, and $\theta_o$ is the horn semi-flare angle. Small
$\Delta$ horns are diffraction-limited and thus have a
frequency-dependent beam size. Large $\Delta$ feeds have a
frequency-independent beam size (the phase error over the feed
aperture effectively washes out coherence).}, the
full-width-half-maximum (FWHM) beam width is
\begin{equation}
\theta_{\rm f}\cong \sin^{-1}\left(\kappa_{\rm h} {\lambda \over
d_{\rm h}}\right),
\label{equation:th}
\end{equation}
where $d_{\rm h}$ is the effective aperture diameter and $\kappa_{\rm
h}\simeq 78^\circ$ is the beam constant for HE$_{11}$ aperture
illumination ($\Delta\simeq 0.1$). To the same order of accuracy, the
main beamwidth is given by
\begin{equation}
\theta_{\rm b} \simeq \kappa_{\rm p} {\lambda \over d_{\rm p}},
\label{equation:tb}
\end{equation}
where $\kappa_{\rm p}\simeq 60^\circ$ is the beam constant for the
primary illumination profile. We define $d_{\rm p}\equiv 2f_{\rm eff}
\tan(\theta_{\rm f})$ as the effective primary illumination
diameter and $f_{\rm eff} \equiv {2f_o/(1+\cos(\theta_{\rm par}))}$
as the distance from the focal point to the center of the off-axis
parabolic section. Solving for the main beamwidth in the limit
$\theta \ll 1$, we find
\begin{equation}
\theta_{\rm b}(\nu) \approx \theta_{\rm b}(\nu_{\rm c})
\left[1-\epsilon+\epsilon \left({\nu_{\rm c} \over \nu} \right)^2 + ... \right],
\label{equation:beam}
\end{equation}
where
\[
\theta_{\rm b} (\nu_{\rm c}) = {\kappa_{\rm p} \over \kappa_{\rm h}}
{d_{\rm h} \over 2 f_{\rm eff}}(1+\epsilon)
\]
and
\[
\epsilon = {1 \over 2}\left( {\kappa_{\rm h} \lambda_{\rm c} \over d_{\rm
h}} \right)^2 \ll 1. 
\]
The center frequency, $\nu_{\rm c}$, is $\sim30.5\,$GHz for ${\rm
K_a}$ and $\sim 41.0\,$GHz for Q-band. The approximate magnitudes of
$\epsilon$, are 0.051, 0.055 and 0.11 for ${\rm SK93K_{\rm a}}$, SK94Q
and SK95Q respectively~\footnote{Equation~\ref{equation:beam} suggests
that in order to reduce the main beam frequency dependence, the
overall size of the optical system should be as large as possible for
a fixed $\theta_{\rm b}$. The residual contributions to the beam's
frequency dependence that arise from non-ideal feed performance and
finite edge illumination should also be considered in this limit.}.

        The telescope is focused by placing the feed phase center at
the primary's focal point. For a corrugated horn the location of the
phase center behind the plane of feed aperture, $t$, for the
fundamental Gaussian mode can be expressed as~(\cite{martin90}),
\begin{equation}
{t \over l_{\rm h}}={\left(\gamma \Delta \right)^2 \over
1+\left(\gamma \Delta \right)^2},
\label{equation:tl}
\end{equation}
where $\gamma\equiv 2\pi\zeta^2=2.603$, $l_{\rm h}$ is the slant
length of the horn, and $\Delta$ is the aperture phase error. Thus,
for a small $\Delta$ horn, the phase center is near the aperture of
the horn, while for large $\Delta$ horns the phase center is near the
apex. The change in the position of the phase center with wavelength
is given by
\begin{equation}
{\partial t \over \partial \lambda} = -{2l_{\rm h} \over
\lambda} {\left( \gamma \Delta \right)^2 \over 1+\left(\gamma
\Delta \right)^2} \left( 1-{\left(\gamma \Delta \right)^2 \over
1+\left( \gamma \Delta \right)^2} \right).
\label{equation:dtdl}
\end{equation}
Designs for frequency-insensitive waist positions exist for $\Delta >
1.8$ and $\Delta < 0.2$. We require that the change in waist position
between 26 and 36 GHz is less than, $\sim 1 \lambda$, the allowed
defocusing error. During the alignment of the optics, the horn
position that maximizes the forward gain in the center channel was
used as the effective focal point. As a result, the
upper and lower frequency channels are not optimally focused but
the average instrument gain is maximized. The best horn position
was insensitive to $< 1.5\lambda$ changes, consistent with
the modeled prediction.

\begin{deluxetable}{llcccc}
\tablewidth{0pc}
\tablecaption{SK Measured and Modeled Beamwidths \label{table:beamwidths}}

\tablehead{
\colhead{}      & 
\colhead{FWHM}  &
\colhead{A1/B1} &
\colhead{A2/B2} &
\colhead{A3/B3} &
\colhead{$\sigma_\theta$} }

\startdata 

${\rm SK93K_a}$ 
        & $\theta_{\rm x}\simeq\theta_{\rm y}$  
                & $1.46^\circ$ 
                & $1.44^\circ$ 
                & $1.41^\circ$ 
                & $\pm 0.02^\circ$ \nl
${\rm SK94K_a}$ 
        & $\theta_{\rm x}\simeq\theta_{\rm y}$  
                & $1.44^\circ$ 
                & $1.42^\circ$ 
                & $1.41^\circ$ 
                & $\pm 0.02^\circ$ \nl
Modeled
        & $<\theta_{\rm b}(\nu_{\rm c})>$
                & $1.47^\circ$
                & $1.44^\circ$ 
                & $1.41^\circ$
                & $\pm 0.03^\circ$ \nl \hline 
${\rm SK94Q}$    
        & $\theta_{\rm x}$      
                & $1.09^\circ$ 
                & $1.08^\circ$ 
                & $1.07^\circ$ 
                & $\pm 0.02^\circ$ \nl
        & $\theta_{\rm y}$      
                & $1.012^\circ$ 
                & $1.004^\circ$ 
                & $0.993^\circ$ 
                & $\pm 0.006^\circ$ \nl
Modeled
        & $<\theta_{\rm b}(\nu_{\rm c})>$                       
                & $1.06^\circ$
                & $1.04^\circ$ 
                & $1.00^\circ$
                & $\pm 0.03^\circ$ \nl \hline 
${\rm SK95Q}$    
        & $\theta_{\rm x}$      
                & $0.471^\circ/0.486^\circ$ 
                & $0.443^\circ/0.461^\circ$ 
                & $0.453^\circ/0.496^\circ$ 
                & $\pm 0.01^\circ$ \nl
        & $\theta_{\rm y}$      
                & $0.567^\circ/0.538^\circ$  
                & $0.525^\circ/0.513^\circ$ 
                & $0.570^\circ/0.591^\circ$ 
                & $\pm 0.01^\circ$ \nl
Modeled
        & $<\theta_{\rm b}(\nu_{\rm c})>$                       
                & $0.513^\circ$
                & $0.505^\circ$ 
                & $0.499^\circ$
                & $\pm 0.02^\circ$ \nl 
\enddata
\tablecomments{The best fit to the measured beam-widths in the vertical
and horizontal planes are denoted by $\theta_{\rm y}$ and $\theta_{\rm
x}$ (The E-plane for the `A' polarization is along $\theta_{\rm y}$
and for `B' it is along $\theta_{\rm x}$.). The modeled
full-width-half-maximum response, $\theta_{\rm b}$, is given for the
center frequency for the design geometry. The uncertainty reflects the
variation in beam-width with polarization and cut plane (E, H, and
diagonal), and the uncertainty in the aperture illumination and
telescope alignment. The discrepancy between the modeled and measured
response for the SK95Q beam results from a 0.4\,cm spacer
(inadvertently omitted during alignment) which vertically shifted the
feed in the focal plane from the design geometry. The observed
increase in the beamwidth along $\theta_{\rm y}$ and decrease along
$\theta_{\rm x}$ is consistent with the phase center offset.}
\end{deluxetable}

\begin{figure}[tbph]
\plotone{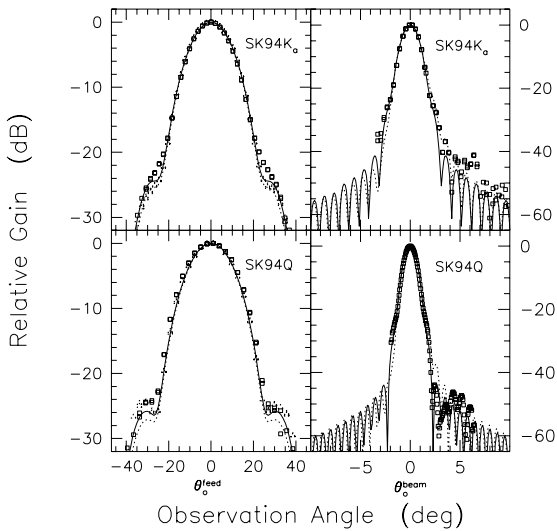}
\caption {The SK94 main beam and feed horn response. The
  measured and 
  theoretical response for the corrugated feed horns used in the ${\rm
    K_a}$ and Q-band systems are presented in the left panels.  The solid
  lines are the E-plane and the dashed lines are the H-plane
  response. The right panels display the measured and computed telescope
  main beam maps.  The theory curves were computed by aperture
  integration of the modeled feed illumination and phase. For clarity,
  only the measured main beam E-plane data for the horizonal
  polarization (channel B) is plotted. All measurements are at the feed
  hybrid frequency $\nu_{\rm x}$ (See
  Table~\protect{\ref{table:telescope}}).} 
\label{plot:bpbeam}
\end{figure}

        The telescope angular response was computed by aperture
integration~(\cite{sletten88}). We model the near field feed amplitude
and phase with the first 20 modes in a Gauss-Laguerre
expansion~(\cite{friberg92}; \cite{tuovinen92}) and integrate the
resulting fields in the chopper plane~\footnote{We ignore the effects
of the feed choke rings and vacuum window in the calculation. For the
parabola, the longitudinal aperture currents are ignored. This
introduces a $\lambda/16$ path-length error for observation angles
\begin{equation}
\theta_o^{\rm beam} > \left( {f_{\rm eff} \lambda/ 2 b^2}
\right)^{1/2} \simeq 10^\circ,
\end{equation}
from boresight, where $2b$ is the primary's major axis. At angles
large compared to this value, the sidelobes are determined by the
details of the primary, chopper, and ground-screen edge illumination,
phase, and fabrication tolerances.}. Figure~\ref{plot:bpbeam} shows a
comparison between the computed and measured beam profiles. A summary
of the main beamwidth for each channel and polarization is given
in~Table~\ref{table:beamwidths}.

        The A and B polarizations were characterized in both the E and
H-plane by illuminating the telescope with a coherent RF source. The
far ground-screen, which has negligible effect on the main beam, was
not present during this measurement.  The measured beam efficiency,
$\eta_{\rm b} \simeq 0.99\pm0.01$, is in agreement with the
model~\footnote{For the SK telescope the aperture diameter is defined
by the parabola edge. The parabola's illumination function encompasses
the first null of the feed pattern over the entire RF bandwidth. In
this limit, the beam efficiency is equal to the feed beam
efficiency. In estimating the total main beam solid angle we integrate
out to $2.5\theta_{\rm FWHM}$. This is approximately the position of
the first null in the response in the limit of a vanishing aperture
phase error. The modeled magnitude includes the estimated ohmic,
diffractive, and depolarization losses intrinsic to the telescope. The
atmospheric transmission efficiency at the site (which can be derived
from Figure~\ref{plot:zenithtemp}), is not included in $\eta_{\rm
b}$.}. After correction for the 80\,m source-to-observation distance
(\cite{silver49}), the measured and computed beamwidths agree to
within $\sim\pm2\%$. The magnitude of this correction is $\ll 1\%,
0.4\%,$ and $12\%$ respectively for the ${\rm SK93K_a}$, SK94Q, and
SK95Q main beamwidths. The computed magnitudes are used to predict
the optical performance. However, for the data analysis we rely on
celestial calibrators to characterize the telescope angular
acceptance.

        The dominant source of cross-polarization in the telescope is
the off-axis parabola. Its computed value is $-24\,$dB
(\cite{rudge78}).  The orthomode provides greater than $-35\,$dB
isolation between the A and B polarizations. The corrugated feed
cross-polar pattern is intrinsically limited by the slot geometry and
throat design to $-40\,$dB (\cite{clarricoats84,dragone77}) over our
observing band. These effects introduce a correlation coefficient of
$\sim0.004$ between the A and B channels which is small compared to
the atmospheric correlations.

        The primary used for SK93/94 is machined out of a solid QC7
aluminum plate. The final surface has a {\rm rms} of $8\mu$m$ < 0.001
\lambda$ (\cite{crone93}). The SK95 mirror (\cite{fixsen95}) was made
out of 6061-T6 aluminum plate in $9$ pieces, a $58\,$cm center section
surrounded by $8$ petals with a $\sim 10 - 20 \mu{\rm m}$ {\rm rms}
surface.

\subsection{The Chopping Flat}

        A large flat aluminum honeycomb plate driven by custom voice
coils sweeps the beam across the sky in azimuth\footnote{The design is
an extension of smaller choppers built at Princeton by M. Dragovan,
J. Peterson, and G. Wright. We benefitted from their insights and
previous work.}. A computer sends the requested position signal to a
PID (Proportional Integral Derivative) control loop similar to that
used by Payne (1976)~\nocite{payne76} and Radford et al.
(1990)~\nocite{radford90}. A high power op-amp drives current through
the coils in response to an error signal (e.g. \cite{kuo91}).

\begin{deluxetable}{lll}
\tablewidth{0pc}
\small
\tablecaption{SK94 Chopper Components \label{table:chopper}}
\tablehead{
\colhead{Assembly}    & 
\colhead{Subassembly}   &
\colhead{Physical Characteristics} \\[.2ex]
\colhead{}            & 
\colhead{Part Number} &  
\colhead{Source} }

\startdata 

Chopping Plate  & Chopper Flat 
                & Thickness = 2.59\,cm (Front/Back Face=0.305\,mm/0.406\,mm) \nl
                & Al Honeycomb 
                & Mass = 5.8\,Kg, I = 0.36\,Kg-m$^2$ \nl            
                && Shape: 3' by 5', approximately ellipsoidal \nl
                && M.C. Gill Co., El Monte, CA  
        \medskip\nl 
                & Flex Pivots & Diameter = 1.27\,cm  \nl            
                & P/N 5016-800
                & Lucas Aerospace, Utica, NY 
        \medskip\nl
                & Angle Encoder 
                & RVDT (rotary variable differential transformer) \nl   
                & P/N RSYN-8/30
                & Lucas-Schaevitz, Pennsauken, NJ 
        \medskip\nl 
                & Servo Drive  
                & Nominal 4\,Hz Triangle: $\pm 70\,$VDC @ 3.4\,A ({\rm rms}) \nl
                & PA-04
                & Apex Microtechnology Corp., Tucson, AZ \medskip \nl \hline
Reaction Bar    & Flex Pivots & Diameter = 1.91\,cm \nl     
                & P/N 5024-400
                & Lucas Aerospace, Utica, NY 
        \medskip\nl 
                & Position Encoder 
                & LVDT (linear variable differential transformer)\nl        
                & Model 503XS-A
                & Lucas-Schaevitz, Pennsauken, NJ 
        \medskip\nl 
                & Servo Drive  
                & Nominal 4\,Hz Triangle: $\pm15$\,VDC @ $1.4$\,A({\rm rms}) \nl
                & PA-12
                & Apex Microtechnology Corp., Tucson, AZ \medskip \nl \hline
Coils/Magnets   & Coil & Resistance = 13.5\,$\Omega$, Inductance = 20\,mH \nl   
                & \# 24 Ga. Cu Wire 
                & Diameter =7.9\,cm, Axis Distance = 29\,cm 
        \medskip\nl     
                & Permanent Magnets & Diameter = 5.08\,cm, Length = 4.76\,cm \nl
                & NdFeB, Grade 35
                & Magnetic Sales and Mfg., Culver City, CA 
        \medskip\nl 
                & Yokes  & Diameter = 12.7\,cm, Length = 12.7\,cm \nl
                & CMI-C Low Carbon Iron
                & Connecticut Metals Incorp., Waterbury, CT \nl 
\enddata
\end{deluxetable}

\begin{figure}[tbph]
\plotone{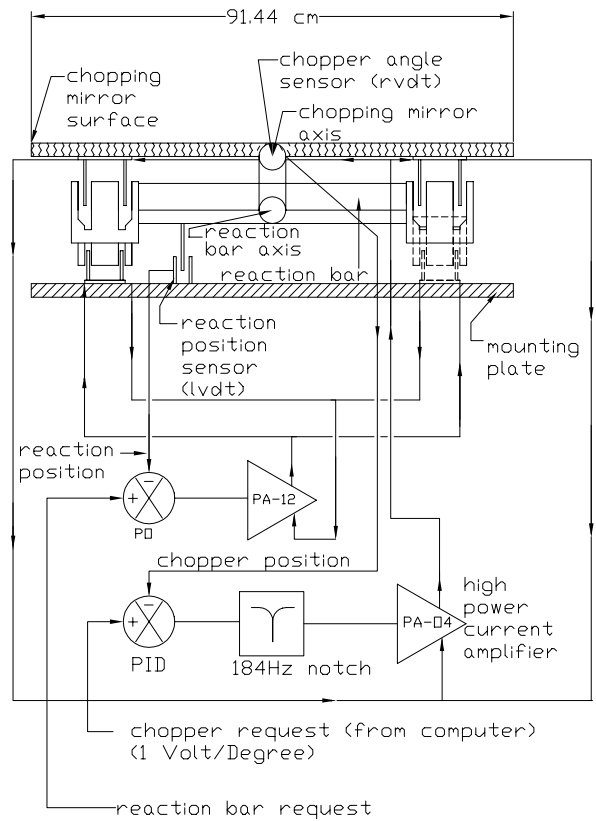}
\caption{Top view of the SK94 chopper. The chopping
  flat and the reaction bar are separately mounted at their centers of
  mass with flex pivots. A notch filter removes frequency components
  from the error signal that are near the 184 Hz bending moment of the
  plate. The angle of the reaction bar is measured with a
  LVDT.} 
\label{plot:chopper} 
\end{figure}

        A schematic of the chopper is shown in
Figure~\ref{plot:chopper}, and the components and dimensions are given
in Tables~\ref{table:telescope} and~\ref{table:chopper}.  Coils are
wound on and epoxied to a Kevlar substrate which is in turn epoxied to
the back of the chopping flat. The rotation axis of the assembly is
adjusted to be at the center of mass, thus the front surface moves a
small amount laterally as the chopper rotates. The coils are
positioned so that an impulse produces minimal force on the
pivot. Permanent magnets are mounted to a reaction bar which is also
pivoted at its center of mass. A second set of `coupling' coils
couples the reaction bar to the frame using a PD control loop. Without
them, the reaction bar hits the chopper mount when the telescope base
is rotated.

        When the coupling coils are turned off, the reaction bar and
chopping plate are well modeled as a driven linear oscillator
(\cite{radford90}). The driving force is $F=2\pi \rho N_{\rm eff} i B$
where $\rho$ is the radius of the coil, $N_{\rm eff}$ ($\approx
120\,$turns) is the number of turns in the magnetic field $B$
($\approx 2000\,$G), and $i$ is the peak current through the
coils. For the SK94 chopper, the force is approximately $30\,$N. Air
resistance is negligible. An good estimate of the chopper throw is
$\phi_{\rm c}({\rm max}) \approx 2rF/I\omega^2$ where $I$ is the
moment of inertia, $r$ is the distance from the center of the plate to
the coils, and $\omega$ is the drive frequency. Using the values in
Table~\ref{table:chopper}, $\phi_{\rm c}({\rm max}) \approx 4.3^\circ$
whereas the actual value was $3.5^\circ$. The measured 10\% to 90\%
response time is $\sim 30~$ms. Though the chopper can be square-wave
chopped; sinusoidal and rounded triangle motions are used for
observing because they result in lower frame vibration and require
less electrical power.

        The azimuthal angle of the flat is measured with respect to
the frame with a RVDT (rotary variable differential transformer)
calibrated to $0.006^{\circ}$ with the absolute encoder that measures
the base motion. The angle is sampled at the data rate. For SK93/94
the chopper angle varied by less than $0.01^{\circ}$ from the
requested position. For SK95, the chopper size was increased from
$91~$cm $\times 123~$cm to $146~$cm $\times 208~$cm. Springs were
attached to the flat and it was driven on resonance at 3\,Hz.  This
led to a large reduction in the required power (SK94/SK93: 200\,Watts,
SK95: 25\,Watts, despite the increase in size) but degraded
positioning. At the worst times, the rms deviation from the requested
position was $0.045^{\circ}$. This has no measurable effect on the
data.

        The magnetic fields of the chopper have been mapped and do not
interfere with the receiver. The DC field within each magnet assembly
is $\approx 2000~$G over a 200\,cm$^3$ volume and drops to less than
5\,G at 20\,cm. The permanent fields are oriented so that the AC
components produced by the coils are in opposite directions near the
receiver. At the receiver, the field at 3.9\,Hz is 0.02\,G.

        The coils heat when the chopper is running and are cooled with
DC fans. Seven sensors monitor the temperature distribution of the
front of the plate. The temperature difference between the center to the
coil position is about 7\,K over 26\,cm and is constant.

        The surface of the chopping plate is not perfectly flat. At
$\sim 2.5\,$cm scales, the deviations are random with a
root-mean-squared amplitude of $<40~\mu$m. This results in a transfer
of $<0.005$ of the relative forward gain into diffuse sidelobes (See
\cite{ruze66}; also, \cite{dragone63} for effects on the wide-angle
antenna response.).  Print-through of the aluminum honeycomb hexagon
pattern produces a periodic array with a $\lambda_{\rm hex} = 0.6~$cm
wavelength and amplitude of $4\,\mu$m. An absolute upper bound on
relative gain loss of $<0.03$ is computed by increasing the Ruze
estimate the total number of scatters, $N \sim (D_{\rm
mirror}/\lambda_{\rm hex})^2$. The measured beam maps do not have
visible grading-lobes every $\lambda/d_{\rm eff} \approx 2^\circ$,
indicating the deviations are not particularly coherent across the
aperture. These surface irregularities have an insignificant effect on
the telescope forward gain and sidelobe response.

\subsection{The Ground-Screens}

        The near and far ground-screens both block the relatively
bright signals from the Earth and Sun and reflect the antenna
sidelobes to cold sky. The far ground-screen size and angle are
designed to shadow the top edge of the chopper baffle for incident
rays greater than the angle at which undiffracted rays are reflected
normal to the panels (see Figure~\ref{plot:setup}). The far
ground-screen is $>17^\circ$ from the main beam, a clearance of $>3.5$
times the parabola width. The ground-screen is fabricated out of
aluminum angle and $1''\times 4'\times 8'$ sheets of metalized housing
insulation (Energy Shield, Atlas Roofing Company, Meridian, MS). The
metal foil covered side of this housing insulation has an antenna
temperature measured at $31.4\,$GHz of $\sim 320\,$mK for a $45^\circ$
incidence angle.  This is a factor of $\sim 1.2$ times the emissivity
computed from the DC conductivity of the aluminum surface. Deviations
in the surface of this material are on the order of 0.2\,cm over a
span of a few centimeters and large scale camber errors are not
uncommon. In effect the sky, which has an order of magnitude lower
antenna temperature than the ground, is being used as a termination
for the sidelobes.

\begin{deluxetable}{lcccc}
\tablewidth{30pc}
\tablecaption{${\rm SK93K_a}$ Offset and Spill Estimates \label{table:syssum}}

\tablehead{     
                & \colhead{$T_{\rm phys}$} & \colhead{Constant} &
                \colhead{Modulation} & \colhead{Offset} \nl &
                \colhead{[K]} & \colhead{[mK]} & \colhead{Source} &
                \colhead{[$\mu$K]} }

\startdata 
$T_{\rm electronic}$        & $-$       & $-$                & C/W & $< 10$ \nl 
$T_{\rm microphonic}$       & $-$       & $-$                & C/W & $< 10$ \nl\hline 
$T_{\rm horn}^{\rm e}$      & $ 12$ & $300$          & $-$ & $-$ \nl 
$T_{\rm window}^{\rm e}$    & $260$ & $20$           & $-$ & $-$ \nl
$T_{\rm parabola}^{\rm e}$  & $260$ & $300$          & $-$ & $-$ \nl 
$T_{\rm plate}^{\rm e}$     & $260$ & $400$          & C   & $< 300$\nl 
$T_{\rm gs}^{\rm e}$        & $260$ & $2$            & C/W & $< 1$ \nl 
$T_{\rm atm}^{\rm e}$       & $260$ & $15000$        & C/W & $< 200$ \nl\hline 
$T_{\rm parabola}^{\rm s}$  & $15$  & $100$          & W   & $< 10$ \nl 
$T_{\rm plate}^{\rm s}$     & $15$  & $20$           & C/W & $< 10$ \nl
$T_{\rm gs}^{\rm s}$        & $260$ & $20$           & C/W & $< 10$ \nl 
\enddata
\tablecomments{The physical temperatures of the emitting object, $T_{\rm
phys}$, are representive of the average magnitude. Temperatures with a
superscript `e' are due to emission. Temperatures in the table with a
superscript `s' are due to antenna spill. `Constant' is the absolute
radiometric brightness. Entries denoted with a `W' and `C' indicate
modulation by the base wobble and the chopper motion respectively. The
offsets are for the three-point data.}
\end{deluxetable}

        The sidelobe response of the telescope, including the
near and far ground-screens, was measured. The structure was
illuminated by a $\theta_{\rm FWHM}=7^\circ$ coherent RF source
placed at eight locations around the perimeter of the
ground-screen and on the roof a building behind the
ground-screen. The source was pointed $\sim 9^\circ$ upward from
horizontal, aimed at the edge of the closest panel, from 12 m
away.  It was modulated by a `hand chopped' AN 73 Eccosorb
(Emerson Cuming, Canton, MA) sheet in front of the transmitting
horn. The measured sidelobe level is less than $-115$\,dB for
rays impinging upon the ground-screen from behind~\footnote{The
sidelobe level is measured relative to power received by the main
beam, $10\log(P_\theta/P_\circ)$.}. For rays illuminating the
front ground screen edges, the response was $\sim -95\,$dB. When
the mirror is chopped, the demodulated signal was measured to
decrease $\sim10~$dB in the E and H-planes relative to the direct
signal from the source. Similar results were obtained for the
Q-band system at 39.5\,GHz. The attenuation provided by
the ground-screen, found by measuring the response to a source directed at
the telescope with and without the ground-screen erected, is greater
than $36\,$dB. Leakage through the lower seams and 
panel glue joints limit the net attenuation. The maximum
`anisotropy' emission from the ground screen is estimated to be
less than $1\mu$K (See Table~\ref{table:syssum}).

        With the exception of the sun, most foreground sources are
more homogeneous than the test source. To the extent signals are not
blocked by the ground-screens and are inhomogeneous after spatial
averaging by the telescope sidelobes, a difference signal is produced.
The ${\rm K_a}$ beam solid angle is $\sim10^{-4}\,$Sr. The sun is a
8200\,K source at $\sim1\,$cm with a solid angle of $6 \times
10^{-5}\,$Sr~(\cite{allen76}). Thus, a sidelobe level of $\sim
-100\,$dB is required to give a response of ${\rm \sim 1\,\mu K}$ from
incident solar radiation. For the Earth, a 300\,K source subtending
$2\pi\,$Sr, a sidelobe of $\sim -130\,$dB is required if a complete
modulation of the incident radiation is assumed. Conservative
analytical estimates indicate that differential contributions to the
antenna temperature from the sidelobes are $< 10\mu$K. A host of
effects enter at this level: signals can result from diffracted
earth/sun-shine, changes in sidelobe illumination with base position,
and gradients in the ground-screen panel temperature and emissivity.

        In 1993, the ground-screen did not geometrically block
the sun when it was directly behind the rear corners (See
Figure~\ref{plot:setup}). Once the problem was noticed, the 
ground-screen was modified. To be conservative, data taken with
the sun illuminating the top of the chopper were
blanked~(\cite{wollack93}). Independent of this precaution, for a
given position on the sky, data recorded during the day and night
are in agreement and the telescope offsets are independent of the
time of day. To accommodate the size of the SK95 optics, the
front edges of the far ground-screen were extended 46\,cm and
rolled with a radius of curvature of $r \simeq 23\,$cm, to reduce
the level of diffracted radiation~(\cite{keller59}). With
these modifications, the net shielding from the front section of
the far ground-screen is similar to the level used in SK93/94.

%
%
%
%
\section{Flux Scale Calibration and Pointing}
\label{section:calibration}
\begin{figure}[tbph]
\plotone{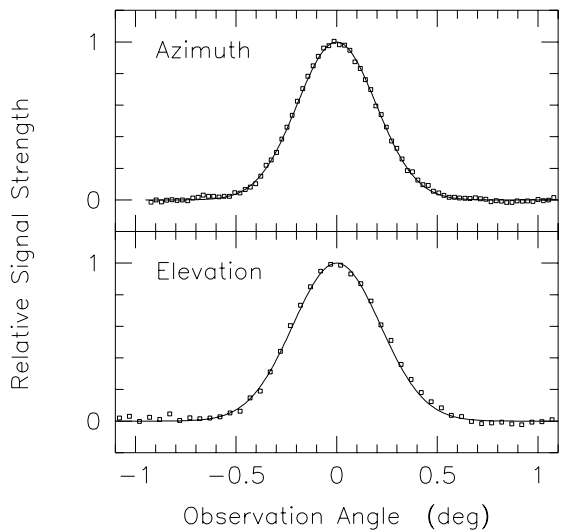}
\caption {The SK95Q beam width determination from Cas\,A.
  Solid lines 
  indicate the best fit Gaussian profiles of the beam. The open squares
  are the radiometer output.} 
\label{plot:casmap}
\end{figure}

        The primary calibrator for both the ${\rm K_a}$ and Q
radiometers is Cassiopeia A~\footnote{3C461, 2321+583 IAU(1950),
$l=111.7^\circ, b=-2.1^\circ$}. Telescope efficiency and atmospheric
attenuation are intrinsically included. The calibration signal is on
the order of $\sim 10\,$mK; thus, receiver non-linearity is not
significant. To determine the beamwidth and pointing, the beam is
swept in azimuth with the chopper while the earth's rotation moves the
source through the beam. The data are reduced with an optimal filter
for a point source to make a two dimensional map of Cas\,A. See
Figure~\ref{plot:casmap}. A two dimensional Gaussian is then fitted to
determine the beamwidth and pointing. A summary of measured beamwidths
is given in Table~\ref{table:beamwidths}.  The pointing is stable for
the duration of an observing trip, attesting to the rigidity and
stability of the telescope platform and footings.

        The flux density scale for Cas\,A was obtained by fitting a
power law to data compiled by~\cite{baars77} from 8.2 GHz to 31.4 GHz
and a measurement by~\cite{mezger86} at 250 GHz.  We find $S_\nu({\rm
Cas\,A}) = (2070\pm 162)\nu^{-0.695\pm 0.029}\,$Jy, where $\nu$ is the
frequency in GHz (epoch 1994). A secular decrease of the form, $\eta =
(0.9903 + 0.003 \log(\nu))^{N_{\rm t}} $ (\cite{baars77}), where
$N_{\rm t}$ is number of years since the measurement, is assumed in
correcting the Cas\,A flux scale to the observing epoch. The measured
point source sensitivity of the telescope, $\Gamma=\lambda^2/2 k
\delta\Omega_{\rm a}$, where $\delta\Omega_{\rm a}$ is the total beam
solid angle~\footnote{The beam solid angle, $4\pi$ divided by the
directivity of the antenna, is related to the main lobe solid angle by
$\delta\Omega_{\rm b}=\eta_{\rm b}\delta\Omega_{\rm a}$, where
$\eta_{\rm b}$ is the beam efficiency.}, is used in assigning the flux
scale. See Table~\ref{table:observe} for the magnitude of
$\Gamma(\nu_{\rm c})$ at the synchrotron-spectrum weighted centroid,
$\nu_{\rm c}$. Once the flux scale is assigned, the data are converted
to the CMB thermodynamic temperature scale by dividing by the
derivative of the Plank function with respect to temperature
$\left({\partial T_{\rm ant} / \partial T_{\rm CMB}}\right)_\nu = {
x^2 e^x / (e^x -1)^2},$ where $x \equiv h \nu / k T_{\rm CMB}$. Using
$T_{\rm CMB} \cong 2.726\,$K~(\cite{mather94,gush90}), the conversion
factors to the {\rm CMB} temperature scale are 1.02 and 1.05 for
27.5\,GHz and 44.1\,GHz respectively.

        The error in the temperature scale is dominated by the
uncertainty in the knowledge of the flux of Cas\,A and its
environment.  Cas\,A passes through an elevation of $52.24^\circ$
twice a day. For Q-band there was $\sim 7\%$ discrepancy and for ${\rm
K_a}$ a $\sim 12\%$ discrepancy between the results of measurements
taken in the morning and those taken in the evening. This effect is
due to 2311+611, a $\sim 25\,$Jy HII region
(\cite{kallas80,fich86,becker91}), which is intercepted by the beam
during the evening runs, but not during the morning runs. The
uncertainty in our measurement of the relative temperature scale of
Cas\,A is $\sim 5\%$, effecting all channels about equally, and the
uncertainty in the spectral index is $\pm 0.1$. The
temperature scale derived from measurements of Cas\,A are within 20\%
of the laboratory sensitivity measurements.  The calibration is also
affected by small errors introduced by inaccuracies in the beam
$\theta_{\rm FWHM}$ $(2\%)$, the phase of the recorded signals with
respect to the optical axis $(<1\%)$, and the finite size of Cas\,A
$(\ll 1\%)$. The combination of all of these errors leads to a
$\pm14\%$ uncertainty ($1~\sigma$) in the {\rm CMB} temperature scale. 
Reduction of this error will require a more accurate determination of the
flux of Cas\,A.

\subsection{Observing Strategy and Beam Synthesis}
\label{section:stratsynth}

        To measure the sky, the beam is swept in azimuth with a
computer commanded pattern, ${\bf x_{\rm a}}$, with frequency $f_{\rm
c}$ and amplitude $\theta_{\rm t}$ on the sky.~\footnote{ The
chopper azimuth, $\phi_{\rm c}$, is related to sky throw by,
$\theta_{\rm t}\simeq 2 \phi_{\rm c} \cos(\theta_{\rm EL})$, where
$\theta_{\rm EL}$ is the beam elevation.} As a second level of
modulation, the base is pointed at ${\bf x}_\circ$, $\theta_{\rm w}$
west of the North Celestial Pole (NCP) for $\sim 20$ seconds and
$\theta_{\rm w}$ east of the NCP for $\sim 20$ seconds in a manner
similar to \cite{timbie90}. The instantaneous beam position is given
by
\begin{equation}
{\bf \hat{x}}^\prime(t) = {\bf x}_\circ(\theta_{\rm w},t) + {\bf x_{\rm
a}}(\theta_{\rm t},t),
\label{equation:cpos}
\end{equation}
where ${\bf x}_\circ$ and ${\bf x}_{\rm a}$ are parallel to the
horizon.  Data taken with ${\bf x}_\circ$ in the east and west base
positions are analyzed independently.  

        The repositioning of the telescope base in azimuth, or
wobbling, allows observation of the same patch of sky approximately
every 12 hours. Thus a real sky signal can be differentiated from a
residual 24 hour diurnal effect in the data set. This symmetry on the
sky is not quite perfect because circles of constant elevation are not
great circles. Nevertheless, the symmetry is good enough to provide
convincing systematic checks.  Due to the geometry of the
ground-screens, a signal produced by the sun is expected to follow a
9\,Hr-15\,Hr-9\,Hr cycle.  If such a signal were present, it would be
stronger in either the east or west and would interchange in base
position from morning to evening. No such signals were observed.

        In order to probe a range of angular scales the beam is
scanned many beamwidths while rapidly sampling. By specifying the
relative weight of each sample in software, an effective antenna
pattern can be synthesized.  For instance, if the samples with the
chopper positioned in the west are weighted with `$-1$' and the east
samples are `$+1$', the beam pattern resembles a classic single
difference. Consider a weighting vector of the form
\begin{equation}
\small
w_i(n) = \sum_{m=1}^{m({\rm max})} a_m \cos (m\, \omega_{\rm c} t_i +
\delta \phi_{\rm s}) + b_m \sin (m \omega_{\rm c} t_i + \delta
\phi_{\rm s}),
\end{equation}
where $\omega_{\rm c} \equiv 2\pi f_{\rm c}$. The time samples
\begin{equation}
t_i = \left(i({\bf \hat{x}}^\prime)-{1/2}\right)/f_{\rm c} N_{\rm s},
\label{equation:timei}
\end{equation}
are evaluated at the midpoint of each integration bin, where $i$ is
the time sample index and $N_{\rm s}$ is the total number of samples
in a chop. A phase difference between the chopper motion and the
recorded data stream, $\delta\phi_{\rm s}=2\pi\,\delta N_{\rm s} \, /
\, N_{\rm s}$, results from the mechanical and electrical responses.

        With our chopper phase convention, the $a_m$ (the signal
phase) are sensitive to the sky signal and the $b_m$ (the quadrature
phase) are sensitive to any instrumental effects. The desired number
and position of the synthesized lobes on the sky are formed from an
appropriate sum of $a_m$~\footnote{The $a_m$ form a complete set of
the sampled spatial frequencies on the sky. The maximum useful $m$ is
set by the intrinsic telescope beam size and the spatial sampling
rate. The beam synthesis allows the window function sidelobe response
to be tailored, allows optimal filtering of atmospheric and
radiometric offset noise, and can facilitate comparison between
different observations (e.g. The center of the synthesized beam
response can be steered in azimuth from the nominal base
position.). In forming a weighting vector with optimal sky
sensitivity, the $b_n$ are set equal to zero; by symmetry they do not
contribute to a celestial signal.}. Physically, the quadrature phase
is the difference between the data taken during the clockwise chopper
scan minus the data taken during the counterclockwise scan. We label a
synthesized beam by the number of lobes, $n=m+1$, the receiver
waveguide band designation, and observing run year. For example, in
${\rm SK93K_a}$, a sinusoidal modulation of the beam was used and
$a_2$ was the only term in the weighting vector. The resulting pattern
qualitatively resembles a `double-difference' or `three-beam-chop' and
we refer to it as the three-point, or 3pt, response.

        As the beam sweeps across the sky, spatial frequencies along
the direction of motion are detected over a range of audio
frequencies. The anti-aliasing filters and integrators (See
Table~\ref{table:observe}) filter our measurement of the angular power
spectrum of the CMB. In 1993, the effect was the largest ($\sim 5\%$)
and was corrected in the frequency domain by a direct multiplication
of the data.  Because the response of the combination of the filters
is well approximated by a Gaussian, the effect may also be corrected
by allowing $\sigma_{||}$ (See Equation~\ref{plot:syngain}) to be a
function of the position in the sweep. In other words, the correction
is made as a convolution in the spatial domain. This was done in 1994
and 1995 to correct a 12\% effect at the highest frequency.

\begin{figure}[tbph]
\plotone{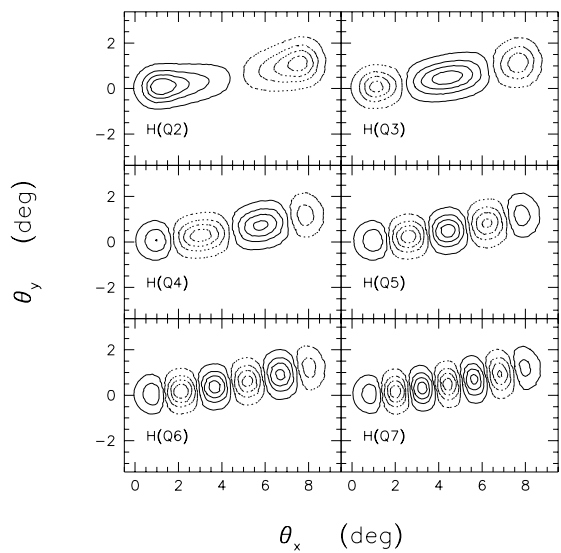}
\caption {Contours of the SK94Q synthesized beam
  pattern. Dashed and solid 
  lines respectively indicate negative and positive beam lobe responses
  during a chop cycle. The telescope is in the east base position. Note,
  lines of constant beam elevation curve upward with respect to lines of
  constant right ascension. The North Celestial Pole (NCP) is at 
  (0,0), $\theta_x$ and $\theta_y$ are defined in
  Table~5.} 
\label{plot:beamcon} 
\end{figure}

\begin{figure}[tbph]
\plotone{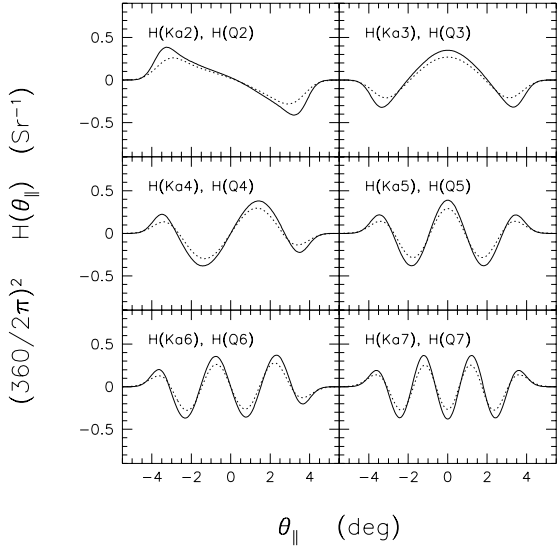}
\caption {Profiles of the SK94 synthesized beam
  patterns. ${\rm 
    SK94K_a}$ and SK94Q are indicated by dashed and solid lines
  respectively. The cut through the synthesized pattern is parallel to
  the direction of the throw. The origin of $\theta_{\|}$ is at the
  chopper center position.} 
\label{plot:beamrsp}
\end{figure}

        The data are normalized so a small change in
temperature of $\delta T$ from a $2.7\,$K blackbody filling the
positive lobes of the synthesized beam gives a detected signal of
$\delta T$.  In other words, the synthesized beams are normalized such
that the integral of the area under the positive lobes of the
effective antenna pattern is equal to one,
\begin{equation}
1 \equiv {1\over2} \int \mid H({\bf \hat{x}},n) \mid {\bf d\hat{x}},
\label{equation:norm}
\end{equation}
where $H$ is the oriented antenna pattern for the $n$-point
response. The oriented antenna pattern is
\begin{equation}
H({\bf \hat{x}},n) \equiv \left\langle{ 
\sum^{\rm N_s}_{i=1} w_i(n) G({\bf \hat{x} - \hat{x}}^\prime(t_i)) 
}\right\rangle,
\end{equation}
where $G$ is the antenna angular response, ${\bf \hat{x}}$ denotes the
arrival direction of photons incident on the telescope, ${\bf
\hat{x}}^\prime$ is the beam position corresponding to sample
$t_i$ as defined in Equation~\ref{equation:cpos}, and the brackets
indicate an average over the pie-like wedge of an
observation~\footnote{The n-point data are averaged for either 30 or
60\,minutes. As a result, the effective synthesized beam pattern for
the final pixel is smeared in right ascension. For atmospheric
investigations, the instantaneous oriented antenna pattern is
used. The difference between normalizing $H$ with the instantaneous
profile and the smeared response is less than 1\%.}. We approximate
the telescope angular response by a Gaussian,
\begin{equation}
\footnotesize
G({\bf \hat{x}}-{\bf \hat{x}}^\prime) 
\simeq {1 \over 2\pi\sigma_\perp \sigma_\parallel} 
\exp{\left(
-{\left| {\bf {x}_\perp - \hat{x^\prime}}(t_i) 
\right|^2 \over 2\sigma_\perp^2} 
-{\left| {\bf {x}_\parallel - \hat{x^\prime}}(t_i)
\right|^2 \over 2 \sigma_\parallel^2}
\right)},
\label{plot:syngain}
\end{equation}
where ${\bf \hat{x}} = {\bf {x}_\perp} + {\bf {x}_\parallel}$, $\sigma
= \theta_{\rm FWHM}/\sqrt{8\ln(2)}$ is the beamwidth, and `$\perp$'
and `$\parallel$' indicate the components perpendicular and parallel
to the direction of the chopper throw. The gain is normalized with an
effective beam solid angle equal to the best fit to the measured
antenna response, $\delta \Omega_{\rm b} \equiv 2\pi\sigma_\perp
\sigma_\parallel$ (the main beam response has a $>0.98$ overlap with a
Gaussian spatial distribution). With these conventions, the signal for
the $j^{\rm th}$ pixel on the sky is
\begin{equation}
\Delta T_j^{(n)} = \int H({\bf \hat{x}},n) T_{\rm b}({\bf \hat{x}})
{\bf d\hat{x}},
\end{equation}
where $T_{\rm b}$ is the celestial brightness distribution. Typical
examples of the resulting beam sensitivities are given in
Figures~\ref{plot:beamcon} and \ref{plot:beamrsp}.

\begin{deluxetable}{lcccc}
\tablewidth{30pc}
\small
\tablecaption{SK Weighting Vector Sensitivity, $\kappa_n$ \label{table:kappa}}
\tablehead{     \colhead{$n$}   
                & \colhead{$\kappa_n({\rm SK93K_a})$}   
                & \colhead{$\kappa_n({\rm SK94K_a})$}   
                & \colhead{$\kappa_n({\rm SK94Q})$}
                & \colhead{$\kappa_n({\rm SK95Q})$}\nl}
\startdata 
        2       & 2.337 & 2.280 & 2.250 & 2.238 \nl
        3       & 3.081 & 2.512 & 2.355 & 2.327 \nl
        4       & .     & 3.010 & 2.573 & 2.451 \nl
        5       & .     & 3.812 & 2.938 & 2.441 \nl
        6       & .     & 5.026 & 3.433 & 2.555 \nl
        7       &       & 6.890 & 4.079 & 2.738 \nl
        8       &       & 9.845 & 4.944 & 2.813 \nl
        9       &       & 14.67 & 6.115 & 3.007 \nl
        10      &       & .     & .     & 3.121 \nl
        11      &       & .     & .     & 3.370 \nl
        12      &       & .     & .     & 3.550 \nl
        13      &       &       &       & 3.859 \nl
        14      &       &       &       & 4.121 \nl
        15      &       &       &       & 4.535 \nl
        16      &       &       &       & 4.867 \nl
        17      &       &       &       & 5.405 \nl
        18      &       &       &       & 5.836 \nl
        19      &       &       &       & 6.619 \nl

\tablecomments{The frequency corresponding to the dominate spectral
component of $w_i$ is given by $f_{\rm obs} \simeq (n-1)\,f_{\rm c}$
where the chopper frequency is $f_{\rm c} \simeq 3.906\,$Hz for ${\rm
SK93K_a}$ and ${\rm SK94K_a/Q}$ and $2.970\,$Hz for SK95Q. }
\enddata
\end{deluxetable}

        The {\rm rms} of the data depends on the weighting vector used. A
measure of this is $\kappa_n$, defined as the ratio of the {\rm rms} of
the weighted data to the {\rm rms} of the raw data,
\begin{equation}
\kappa_n=\left({N_s \sum^{N_s}_{i=1}{w_i(n)^2}}\right)^{1/2}.
\end{equation}
Table~\ref{table:kappa} gives $\kappa_n$ for all the weighting
vectors. The overall sensitivity decreases ($\kappa_n$
increases) as the synthesized beam spacing approaches the intrinsic
telescope beamwidth. In practice, $\sim 3\,$ time samples per beam
width are required in order to avoid a significant loss of resolution.

        The sensitivity of the telescope for a particular weighting
scheme is $\kappa_n S_{\nu}(f_{\rm obs})$, where $f_{\rm obs}$ is the
effective sampling frequency associated with the weighting vector (See
Table~\ref{table:kappa}) and $S_{\nu}$ is the power spectrum of the
receiver while staring at the sky (See
Figure~\ref{plot:tpfft}). Although one looses sensitivity for high
$n$, this is partially compensated by the corresponding reduction
in instrumental $1/f$ noise with increasing frequency.

%
%

%
\section{Radiometric Offsets}
\label{section:offset}

        We use the following model to identify contributions to
the offset:
\begin{eqnarray}
\small
T_{\rm ant} & \cong & (1-\varepsilon_o) 
\left[ \eta_{\rm b} \left( T_{\rm sky} + T_{\rm atm} \right) +
\left( 1-\eta_{\rm b} \right) T_{\rm spill} \right] \nonumber \\
 & & \mbox{} + \varepsilon_o \eta_{\rm b} T_{\rm plate},
\end{eqnarray}
where $\eta_{\rm b} \approx 1$ is the beam efficiency, $\varepsilon_o
\ll 1$ is the emissivity of the aluminum chopping plate, $T_{\rm sky}$
is the brightness temperature of the sky measured from the surface of
the Earth, $T_{\rm atm}$ is the atmospheric brightness temperature,
$T_{\rm plate}$ is the physical temperature of the chopping plate, and
$T_{\rm spill}$ is the termination temperature of the sidelobes.  When
the beam is moved on the sky, the change in antenna temperature is
\begin{eqnarray}
\delta T_{\rm ant} & \simeq &
\delta\eta_{\rm b} \left( T_{\rm sky} + T_{\rm atm} - T_{\rm spill}
\right) \nonumber \\
& & + \delta\varepsilon_o \eta_{\rm b} T_{\rm plate}
 +\eta_{\rm b} \left(\delta T_{\rm atm} + \varepsilon_o \delta T_{\rm
plate} \right) \nonumber \\
& & + (1-\eta_{\rm b})\delta T_{\rm spill},
\end{eqnarray}
where contributions are grouped by changes in coupling efficiency,
emission, and termination temperature\footnote{Modulation of the
receiver input match can also produce synchronous offsets. The beam
switch in the SK telescope is produced by both the chopper and the
base motions. Due to the off-axis telescope geometry and relatively
wide channel bandwidth, the receiver noise temperature and amplitude
response are essentially unmodulated by switching.}. For simplicity,
the sky is assumed to be uniform and only the dominate contributions
to the offset are retained. Offsets due to modulated spill-over
radiation are controlled by baffle geometry and illumination
level. Under-illuminated of the optical elements results in $\delta
\eta_{\rm b} \approx 0$. The offsets from atmospheric emission are
minimized through telescope alignment and by rapidly differencing
regions with the same temperature. Under stable atmospheric
conditions, the dominant contributions are due to modulation of the
antenna spill and plate emissivity.

\subsection{The Chopping Plate Emission Offset}

\begin{deluxetable}{lcccc}
\tablewidth{30pc}
\tablecaption{Measured and Estimated Average Radiometric Offsets \label{table:offsets}}

\tablehead{     
                  \colhead{Component}                   
                & \colhead{${\rm SK93K_a}$}             
                & \colhead{${\rm SK94K_a}$}             
                & \colhead{${\rm SK94Q}$}               
                & \colhead{${\rm SK95Q}$}\nl
                  \colhead{}            
                & \colhead{A/B}                 
                & \colhead{A/B}                 
                & \colhead{A/B}                 
                & \colhead{A/B}\nl 
                  \colhead{}            
                & \colhead{$[\mu{\rm K}]$}              
                & \colhead{$[\mu{\rm K}]$}              
                & \colhead{$[\mu{\rm K}]$}              
                & \colhead{$[\mu{\rm K}]$}\nl}

\startdata 

$\Delta T_{\rm plate}$  
        & $+50/-190$
        & $+120/-430$
        & $+130/-500$
        & $+150/-570$ \nl
$\Delta T_{\rm atm}$    
        & $-150$
        & $-5$
        & $-10$
        & $-10$ \nl
$\Delta T_{\rm spill}$
        & $-10$
        & $-10$
        & $-10$
        & $\sim-500$ \nl \hline
$\Delta T_{\rm theory}$
        & $-110/-350$
        & $+105/-445$
        & $+110/-520$
        & $\cdots$ \nl
$\Delta T_{\rm observed}$
        & $-60/-360$
        & $+117/-550$
        & $\ldots/-695$
        & $-616/-2182$  
\enddata
\tablecomments{The three-point offsets are computed from the measured
plate emissivity, telescope alignment, and spill. $\Delta T_{\rm
spill}$ is an order of magnitude estimate based on the telescope
configuration. The theoretical offset is estimated from $\Delta T_{\rm
theory} = \Delta T_{\rm plate} + \Delta T_{\rm atm} + \Delta T_{\rm
spill}$ using the measured telescope parameters. The offset for the
SK95Q three-point is larger than the simple model used here would
predict. We believe the excess offset is due to a 0.04\,cm wide seam
in one side of the chopper which moves laterally with respect to the
center of the beam during a chop cycle.}
\end{deluxetable} 

\begin{figure}[tbph]
\plotone{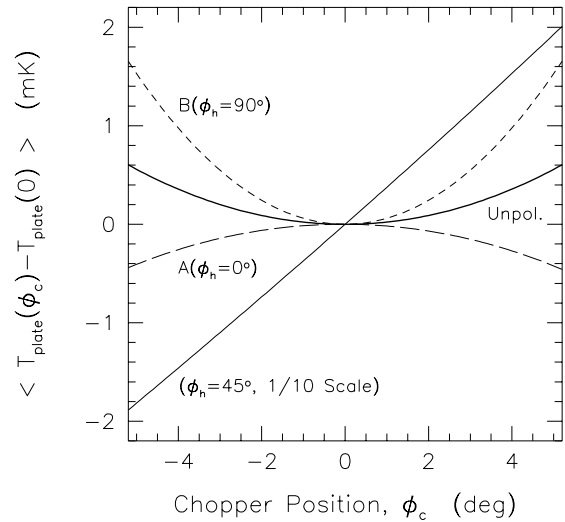}
\caption {The antenna temperature due to plate emission
  as a 
  function of chopper azimuthal position. In the field, feed
  polarization `A' (long dashes) is vertical and `B' (short dashes)
  is horizontal. For this
  calculation, the observation angle is $\theta_{\rm z}=37.85^\circ$ and
  the radial extent of the plate is taken as infinite. A physical
  temperature of $T_{\rm phy}=300\,$K and an emissivity of
  $\varepsilon_o = 9\times 10^{-4}$ are assumed. The solid line is 1/10
  of the computed magnitude for the feed polarization with $\phi_{\rm
    h}=45^\circ$. The response of an ideal unpolarized
  detector is indicated by the solid bold line. The average emission for
  a fixed incidence angle, $260 < \left< T_{\rm plate} \right>
  <430\,$mK, is a function of the feed polarization orientation. To
  facilitate comparison, this constant term has been subtracted from the
  computed response. Calculated and measured magnitudes are given in
Table~\protect{\ref{table:offsets}}.} 
\label{plot:chopoff}
\end{figure}

        The process of scanning with the chopper produces a
synchronous modulation of the plate emissivity when viewed from the
feed. As the angle of incidence is varied, the magnitude of the
chopper's surface brightness temperature changes. For good conductor
with skin depth $\delta = \left(1 / \mu_o \pi \nu_o \sigma
\right)^{1/2}$, where $\sigma$ is the conductivity of the metal
surface and $\nu_o$ is the observing frequency, the parallel and
perpendicular emissivities are
\begin{equation}
\varepsilon_{\|} \simeq {\varepsilon_o / \cos(\theta_{i})},
\label{epar}
\end{equation}
with $\delta/\lambda_o \ll \cos(\theta_{i})$, and
\begin{equation}
\varepsilon_{\perp} \simeq \varepsilon_o \cos(\theta_{i}),
\label{eper}
\end{equation}
with $\delta/\lambda_o \ll 1$. The incident angle of the radiation
is given by
\begin{equation}
\theta_i(\phi_{\rm c}) = \cos^{-1} \left(-{ \bf \hat{k}}_i \cdot {\bf
\hat{n}}(\phi_{\rm c}) \right),
\end{equation} 
where ${ \bf \hat{k}}_i$ is the propagation vector and ${\bf \hat{n}}$
is the plate normal vector in the detector
frame. (e.g. \cite{landau60}; \cite{rytov78}). The emissivity of a
6061-T6 aluminum sheet at $31.4\,$GHz is $\varepsilon_{o} \equiv 4\pi
{\delta / \lambda_o} \approx 9\times10^{-4}$. We normalized
Equations~\ref{epar} and~\ref{eper} to the DC conductivity of the
surface~\footnote{Due to the differences in the DC conductivity of
aluminum, the relative emissivity varies by a factor upto $\sim 1.5$
depending upon alloy type (e.g. \cite{crc82}). Surface treatment and
finish can increase the RF emissivity by an additional factor between
$1.1$ and $1.5$ in typical microwave components.}. The brightness
temperature of the plate is
\begin{equation}
T_{\rm plate} \simeq T_{\rm phy} \left[
\varepsilon_{\|}(\theta_{i})\left|{\bf
\tilde{E}}_{\|}(\theta_{i})\right|^2 +
\varepsilon_{\perp}(\theta_{i})\left|{\bf
\tilde{E}}_{\perp}(\theta_{i})\right|^2 \right],
\label{tplate}
\end{equation} 
where $T_{\rm phy}$ is the physical temperature of the plate, and
${\bf \tilde{E}}_{\|}$ and ${\bf \tilde{E}}_{\perp}$ are the
perpendicular and parallel electric field projections onto the
chopping flat normal. The resulting emission can be expressed as
\begin{equation}
\Delta T^{(n)}_{\rm plate} \simeq {\int H({\bf \hat{x}},n) T_{\rm
plate}({\bf \hat{x}}) {\bf d\hat{x}}},
\end{equation}
where $H$ is the antenna pattern of the synthesized beam (See
Section~\ref{section:calibration}). Since the chopping plate
essentially fills the entire field of view, the emission offset is
only a function the chopper position angle and the weighting
vector. The antenna temperature as a function of the physical chopper
position angle with $w_i=1$, is given in
Figure~\ref{plot:chopoff}. For polarization $\phi_{\rm h}=0^\circ$ (E
vector ${\perp}$ to the chopper axis) and $90^\circ$, the
anti-symmetric component of the offset is eliminated. With $\phi_{\rm
h}=\pm 45^\circ$ the symmetric term is minimized. For an unpolarized
detector, the anti-symmetric component is eliminated by the average
over feed polarization angle. The measured and computed three-point
offset for each radiometer is summarized in Table~\ref{table:offsets}.

        With $\phi_{\rm h}=\pm 45^\circ$, the two-point offset
was measured to be $\delta \Delta T_{\rm plate}/ \delta \phi_{\rm
c}\approx \pm (3.3 \pm 0.3)\,{\rm mK\,deg^{-1}}$. The predicted
magnitude is $\sim 3.5\,$mK${\rm \,deg^{-1}}$ for the aluminum
chopping plate surface. In the vertical \& horizontal feed
polarization configuration ($\phi_{\rm h}=0^\circ$ and
$90^\circ$) a stainless steel sheet was rigidly attached to the
chopper surface. The measured offset increased by a factor of
$\sim 7$ consistent with the increase in emissivity.

        Changes in the plate temperature influence the offset
magnitude. The chopping flat temperature is monitored at 7
locations. While taking astrophysical data, the average chopping plate
temperature varied by as much as $30\,$K. This results in a maximal
$\pm \sim 5\%$ variation in the offset magnitude as a function of the
ambient temperature. There is a small correlation between the chopper
temperature and the offset. However, changes in ambient temperature
can effect the offset through other mechanisms (See
Section~\ref{section:offvar}) so one cannot conclude that the
correlation is causal. Removal of the correlated signal has negligible
effect on the reported results.

\subsection{The Chopping Plate Spill Offset}

        As the plate moves, the chopper edge diffraction changes. This
effect is minimized by under-illuminating the flat.  By evaluating the
directivity in the Gaussian optics approximation we obtain an estimate
of the chopper edge illumination (\cite{martin90}; also see,
\cite{murphy93}); for the SK93 optics it is $-46\,$dB relative to the
near field main beam, consistent with measurement. In addition, a
stationary baffle that extends $15\,$cm past the chopper edge shields
the chopper from behind.  The computed illumination level at this edge
is more than $-65\,$dB lower than the main beam in the chopper
plane. For the ${\rm SK93}$ and SK94 optical designs, the offset due
to modulated spill is immeasurably small. For SK95, the edge
illumination was increased to produce a narrower beam. This
contributed to the larger 3pt offset that year.

\subsection{The Atmospheric Offset}

\begin{figure}[tbph]
\plotone{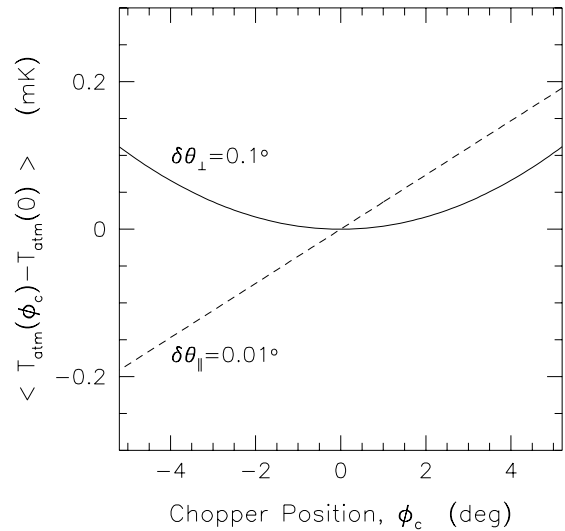}
\caption {The atmospheric antenna temperature due to
  misalignment of the 
  chopper rotational axis as a function of chopper azimuthal position. A
  zenith temperature of $T_{\rm z}=10\,$K is assumed.  Typical axis
  misalignment errors, $\delta \theta_\perp=0.1^\circ$ and $\delta
  \theta_\|=0.01^\circ$ are indicated by solid and dashed lines
  respectively. The three-point offset (inner minus outer lobes), is
  negative for a positive elevation error.} 
\label{plot:atmsoff}
\end{figure}

        The chopping plate axis is aligned with the
vertical in order to minimize the atmospheric offset, $\Delta T_{\rm
sky}$. If the plate axis is rotated in the plane of the plate,
$\delta\theta_\|$, a two-point response will be
generated. A base or chopper throw leveling error would induce this
offset. A vertical axis error due to a rotation perpendicular to the
plane of the plate, $\delta\theta_\perp$, generates a three-point
response. A typical example of this is an elevation pointing
error. We model atmospheric brightness temperature as
\begin{equation}
T_{\rm atm} = \langle T_{\rm atm} \rangle + {\partial T_{\rm atm}
\over \partial \psi}\left[\delta \psi_\|+\delta \psi_\perp\right],
\label{equation:tatmos}
\end{equation}
where $\delta \psi \ll 1$ is the variation in angle due to chopper
misalignment from a fixed zenith angle. We assume a gradient in
the atmospheric temperature of the form
\begin{equation}
{\partial T_{\rm atm} \over \partial \psi} \simeq T_{\rm z}
\tan(\theta_{\rm z})\sec(\theta_{\rm z}),
\end{equation}
where $T_{\rm z}$ is the zenith temperature. The variation in the sky
signal due to a small rotation about the axis parallel to the
chopper normal is approximated by
\begin{equation}
\delta \psi_{\|} \simeq 2 \delta \theta_\| \sin(\theta_{\rm
z}) \sin(\phi_{\rm c}),
\end{equation}
where $\delta\theta_\| \ll 1$. The leading factor of 2 results from
the reflection of the beam off the plate. Similarly, for a small
rotation of the chopper vertical axis about the perpendicular to the
plate normal
\begin{equation}
\delta \psi_{\perp} \simeq -2 \delta \theta_\perp
\cos(\theta_{\rm z})\cos(\phi_{\rm c}),
\end{equation}
for $\delta\theta_\perp \ll 1$. From Equation~\ref{equation:tatmos}, 
the resulting atmospheric offset is
\begin{equation}
\Delta T^{(n)}_{\rm atm} \simeq {\int H({\bf \hat{x}},n) T_{\rm
atm}({\bf \hat{x}}) {\bf d\hat{x}}}.
\end{equation}
In Figure~\ref{plot:atmsoff}, the atmospheric brightness temperature
as a function of chopper position is plotted for typical chopper
alignment errors. In practice, temporal variations in the horizontal
atmospheric temperature profile can mask the effects of the parallel
axis misalignment on short time scales.

\subsection{Limits to Offset Stability}
\label{section:offvar}

        The stability of the offset is as important as its magnitude.
For example, {\sl COBE/DMR} had offsets on the order of hundreds of
milli-Kelvin (\cite{kogut92}) but the extreme stability of the space
environment allowed it to produce the highest quality data set to
date. Stability is especially important for the SK experiment because the
sky is re-observed only every 12 hours. Possible causes for a drift
include changes in the chopping plate temperature profile, thermal
contraction of the mounts, icing of the optics, and changes in the
system gain from thermal variations. The largest drift is $7~\mu$K/day
in the SK95 three-point data, more typically the drift is
$<4~\mu$K/day.

\begin{figure}[tbph]
\plotone{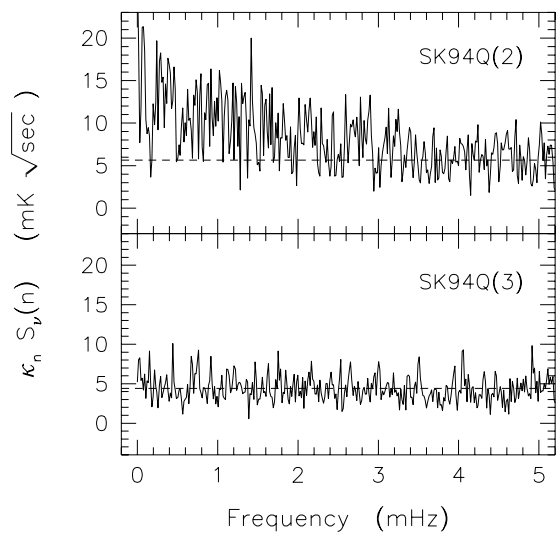}
\caption{The power spectrum of the weighted data (SK94Q:
  $\zeta < 2.5 
  \,{\rm mK\,sec^{1\over2}\,deg^{-1}}$). The dashed lines indicate the
  detector sensitivity. The 2pt data, taken at 4\,Hz and labeled
  SK94Q(2), have a substantial atmospheric contribution, while the
  3pt (and higher) data are detector noise limited. The offset
  for the channel was subtracted before taking the Fourier
  transform.} 
\label{plot:fft} 
\end{figure}

        There may be drifts in the data on faster time scales but the
receivers are not sensitive enough to detect them.  In 12 hours, the
3pt SK94Q data (Figure~\ref{plot:fft}) can only measure
$20~\mu$K with a signal-to-noise of one. Even when data from multiple
channels are combined, the noise is not reduced by more than two. If
there are short time scale drifts, then they must average out because
we observe the same signal on the sky in subsections of the data and
in multiple years. The problem of offset removal is addressed in
\cite{netterfield96}.

%
%
%
%
\section{Atmospheric Noise and Data Editing}
\label{section:atmos}

        During January through March, Saskatoon is relatively dry,
clear, and cold. Precipitable water vapor has a mean value of 5\,mm
and drops to $\le 2\,$mm on the coldest winter
days~(\cite{jones93}). From ground-based measurements of the wind
speed/direction, temperature, pressure and relative humidity at the
site, the following general trends were noted: `good data' are
typically obtained during periods with little or no wind
($<5\,$knots), temperatures less than $-5^\circ\,$C, atmospheric
pressure greater than 710 Torr, stable weather systems, and low
relative humidity.  However, at times the effects of atmospheric
turbulence were seen in the data despite clear skies and favorable
weather station readings, suggesting that the offending fluctuations
were far removed from the telescope. In general, the presence of high
altitude clouds or ice fog did not increase the atmospheric noise.

\begin{figure}[tbph]
\plotone{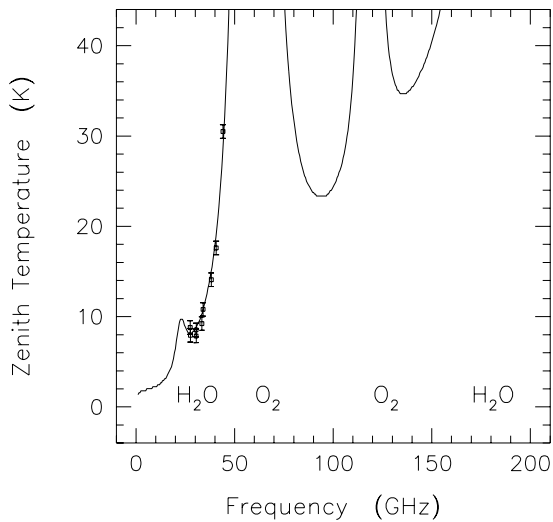}
\caption {The atmospheric zenith temperature in
  Saskatoon, SK on a 
  typical calm winter day. The units are antenna temperature and the
  error bars reflect the uncertainly in the calibration scale. The solid
  curve is the modeled radiometric brightness
  temperature.} 
\label{plot:zenithtemp}
\end{figure}

        The atmospheric zenith temperature is measured by tilting the
chopping plate $\pm 5^\circ$ and measuring the total power at the
detector.  Using the Van Vleck-Weisskopf line shape, we compute the
theoretical atmospheric zenith temperature assuming a
$\sec(\theta_{\rm z})$ dependence for the column
depth~(\cite{liebe85}; \cite{danese89}). The results are presented in
Figure~\ref{plot:zenithtemp}. Data suitable for {\rm CMB} analysis lay
within a $\sim 4\,$K atmospheric temperature window above the minimum
recorded magnitude at the site.

        In addition to the flat spectral component from atmospheric
thermal emission, there is also a `$1/f$' component originating from
atmospheric turbulence (\cite{tatarski61}, \cite{andreani90},
\cite{church95}). There also appear to be other mechanisms that
enter on longer time scales.
The spatial power spectrum of the emission is
dominated by thermal gradients spanning $>4^{\circ}$ as determined
from the 2pt data. These gradients have been observed to persist
for up to an hour. The 3pt data, which are sensitive to the
curvature rather than the gradient, are typically have a factor
of eight less noise.

\begin{figure}[tbph]
\plotone{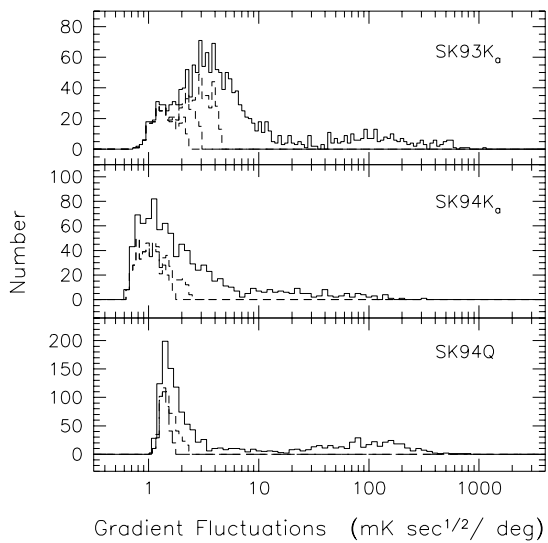}
\caption {The distribution of fluctuations for 20\,minute
  averages of the 
  two-point data as a function of cut level: ${\rm SK93K_a}$ with the
  $27.5\,$GHz channel, $\zeta=2.2, 3.0, 4.5$, and all data; ${\rm
    SK94K_a}, \zeta=1.7, 2.5,$ and all data; and ${\rm SK94Q}$ with the
  $38.5\,$GHz channel, $\zeta=1.7, 2.5,$ and all data. Horizontal units
  are ${\rm mK\,sec^{1\over2}\,deg^{-1}}$ (antenna temperature). The
  receiver noise floors for 1993 and 1994 ${\rm K_a}$ were the same;
  however, with the increased chopper throw in 1994, the sensitivity to
  atmospheric fluctuations increased. To convert to the equivalent ${\rm
    NET_{\rm atmos}}$ multiply by, $\theta_{\rm eff}/\kappa_2=1.68^\circ$,
  for ${\rm SK93K_a}$, $\theta_{\rm eff}/\kappa_2=2.25^\circ$, for ${\rm
    SK94K_a}$, and for SK94Q, $\theta_{\rm eff}/\kappa_2=2.37^\circ$. Note
  that the receiver noise contribution is included in
  $\zeta$.} 
\label{plot:atmnoise}
\end{figure}

        The atmospheric cut for selecting astrophysical data
(3pt and higher) is based on the stability of the
statistically independent 2pt data.  This is done by
evaluating the 2pt mean deviation, $\eta$, of the 8 second
averages for a 20\,minute segment of data. The mean deviation is
less sensitive than the standard deviation to outliers. To
compare with other experiments, the cut levels are converted to
units of ${\rm mK\,sec^{1\over 2}\,deg^{-1}}$ by multiplying
$\eta$, by the sensitivity of the synthesized beam pattern to a
$1\,{\rm mK\,deg^{-1}}$ horizontal atmospheric gradient,
\begin{equation}
\zeta \equiv {\kappa_2 \over \theta_{\rm eff}}  {\rm NET_{\rm atm}}
\approx \sqrt{\pi \over 2}{\eta \sqrt{\tau} \over \theta_{\rm eff}},
\end{equation}
where ${\rm NET_{\rm atm}}$ is the equivalent noise temperature of the
cut level, $\tau$ is the integration time, and $\theta_{\rm eff}={\int
\theta_{\|}({\bf \hat{x}}) H({\bf \hat{x}},2) {\bf d\hat{x}}}$ is the
2pt effective beam separation angle. This conversion assumes
Gaussian fluctuations which is not necessarily valid.
Figure~\ref{plot:atmnoise} shows the distribution of the spatial
gradient fluctuations, $\zeta$, at several cut levels.

        The distributions from ${\rm K_a}$ and Q have some qualitative
differences. The lower cut off on both distributions is due to the
system noise, which is higher in Q than ${\rm K_a}$. The second hump
located in the vicinity of $\sim 200\,{\rm mK\,sec^{1\over
2}\,deg^{-1}}$ is due to data acquired during a two week period of
very poor weather in January 1994. Similarly, the bump in SK93 results
from data taken during the spring thaw in March of 1993. During these
periods the zenith temperature is unstable and the atmospheric water
content relatively high. 

\begin{figure}[tbph]
\plotone{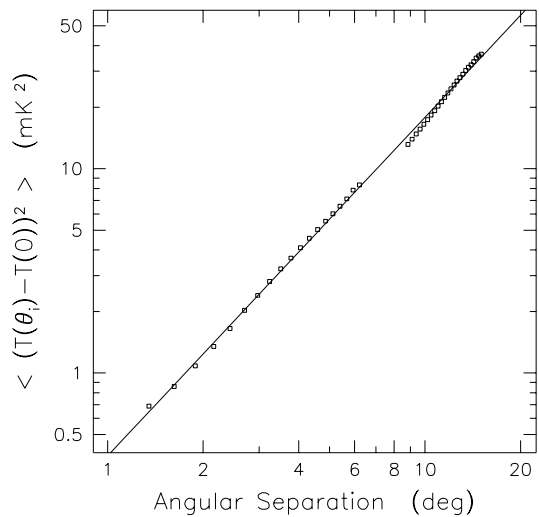}
\caption{The atmospheric noise structure function. The boxes
  are a 20\,min average of the structure function (SK94Q: $\zeta < 2.5
  \,{\rm mK\,sec^{1\over2}\,deg^{-1}}$). The solid line (\protect{$\log
    \langle \left( T(\theta_i-T(0)) \right)^2 \rangle = 1.66 \log(\theta_i)
    -0.41$}) is a fit to the data.} 
\label{plot:structure} 
\end{figure}

        For roughly 25\% of any campaign, the receivers are detector
noise dominated (\cite{netterfield96}). As an example of the
atmospheric stability, the power spectrum for an $18\,$Hr contiguous
stretch of good data ($\zeta < 2.5\,{\rm mK \, sec^{1 \over 2} \,
deg^{-1}}$) is shown in Figure~\ref{plot:fft}. The two-point response,
SK94Q(2), has a $1/f$ knee of $\sim 1\,$mHz and the three-point,
SK94Q(3), is featureless. For $n>2$ the responses are qualitatively
similar to SK94Q(3). In Figure~\ref{plot:structure}, the structure
function, $D(\theta_i) \equiv \langle \left( {T(\theta_i) - T(0)}
\right)^2 \rangle$ (See \cite{tatarski61}, \cite{church95}), is
computed for a 20\,minute average of the data. For beam separations
less than $7^\circ$, the average of the east and west data sets is
used, while for angles greater than $7^\circ$, the east and west data
are combined into a long scan across the sky. The absence of evidence
of saturation in $D(\theta_i)$ suggests this component of the
atmospheric noise results from angular scales greater than the
base wobble angle.

%
%
%
%
\section{Summary}
\label{section:conclusion}

The instrument uses a HEMT-based total-power receiver at the focus of an
off-axis parabolic primary. The beam is swept many beamwidths on the sky
with a large under-illuminated movable flat. The clear optical path,
polarization orientation, and mechanical stability result in small and
stable radiometric offsets. To provide spectral discrimination, each
receiver observes in three frequency bands. The radiometer outputs are
sampled rapidly compared to the beam scan rate.  This is possible
because of the fast response of the detectors.  By weighting each sample
in software, a set of synthesized beams is constructed. This allows one
to simultaneously probe a range of angular scales, optimize spatial
frequency coverage while minimizing atmospheric contamination, and
synthesize beam patterns for a direct comparison with other experiments.

The data from this instrument are of generally high quality and we are
not aware of any instrument-based systematic effects that could
compromise them. The data selection criteria are not based on the data
used for the astrophysical analysis but rather on an independent
weighting scheme. Also, the symmetries in the observing strategy and the
three years of observations provide many internal consistency checks.
The dominant contribution to the $14\%$ calibration uncertainty is from
the inaccuracy in our knowledge of our calibration source, Cas-A.
Future observations using the techniques described here will enhance our
confidence in the data and lead to a better understanding of the
anisotropy in the CMB.

%
%

\section{Acknowledgments}

We gratefully acknowledge the contribution and support of many people
associated with this project. We are indebted to Marian Pospieszalski
and Mike Balister of NRAO for providing the {\rm HEMT}
amplifiers. George Sofko, Mike McKibben at the University of
Saskatchewan's Institute for Space and Atmospheric Studies, and Larry
Snodgrass at the Canadian SRC provided valuable assistance in the
field. We also thank Chris Barnes, Carrie Brown, Weihsueh Chiu, Randi
Cohen, Peter Csatorday, Cathy Cukras, Peter Kalmus, John Kulvicki,
Wendy Lane, Young Lee, Glen Monnelly, Rob Simcoe, Jeno Sokoloski,
Naser Quershi, Peter Wolanin and the Princeton University shop for
their help in constructing the apparatus. Discussions with Sarah
Church, Brian Crone, Dale Fixsen, Tom Herbig, John Ruhl, and Suzanne
Staggs are greatly appreciated.

This research was supported by NSF grant PH89-21378, NASA grants
NAGW-2801 and NAGW-1482, and a Packard Fellowship, a Research
Corporation Award, and an NSF NYI grant to L. Page.


\clearpage

\end{document}